\documentclass[sigconf]{acmart}

\usepackage{graphicx}
\usepackage{multirow}
\usepackage{tabularx}
\newcolumntype{C}[1]{>{\centering\let\newline\\\arraybackslash\hspace{0pt}}m{#1}}
\usepackage[caption=false,font=footnotesize]{subfig}

\DeclareMathOperator*{\argmax}{arg\,max}

\usepackage{wasysym}
\newcommand{\rt}[1]{\textcolor{red}{#1}}

\usepackage{xspace}
\newcommand{\sys}{NNoculation\xspace}

\AtBeginDocument{%
  \providecommand\BibTeX{{%
    \normalfont B\kern-0.5em{\scshape i\kern-0.25em b}\kern-0.8em\TeX}}}

\copyrightyear{2021}   
\acmYear{2021}   
\setcopyright{acmcopyright}
\acmConference[AISec'21]{Proceedings of the 14th ACM Workshop on Artificial Intelligence and Security}{November 15, 2021}{Virtual Event, Republic of Korea}
\acmBooktitle{Proceedings of the 14th ACM Workshop on Artificial Intelligence and Security (AISec '21), November 15, 2021, Virtual Event, Republic of Korea}
\acmPrice{15.00}
\acmDOI{10.1145/3474369.3486874} 
\acmISBN{978-1-4503-8657-9/21/11}

\begin{document}

\title{NNoculation: Catching BadNets in the Wild}

\author{Akshaj Kumar Veldanda}
\affiliation{%
  \institution{New York University}
  \city{New York}
  \country{USA}
}
\email{akv275@nyu.edu}

\author{Kang Liu}
\affiliation{%
  \institution{Huazhong University of Science and Technology}
  \city{Wuhan}
  \country{China}
}
\email{kangliu@hust.edu.cn}

\author{Benjamin Tan}
\affiliation{%
  \institution{University of Calgary}
  \city{Calgary}
  \country{Canada}
}
\email{benjamin.tan1@ucalgary.ca}

\author{Prashanth Krishnamurthy}
\affiliation{%
  \institution{New York University}
  \city{New York}
  \country{USA}
}
\email{prashanth.krishnamurthy@nyu.edu}

\author{Farshad Khorrami}
\affiliation{%
  \institution{New York University}
  \city{New York}
  \country{USA}
}
\email{khorrami@nyu.edu}

\author{Ramesh Karri}
\affiliation{%
  \institution{New York University}
  \city{New York}
  \country{USA}
}
\email{rkarri@nyu.edu}

\author{Brendan Dolan-Gavitt}
\affiliation{%
  \institution{New York University}
  \city{New York}
  \country{USA}
}
\email{brendandg@nyu.edu}

\author{Siddharth Garg}
\affiliation{%
  \institution{New York University}
  \city{New York}
  \country{USA}
}
\email{sg175@nyu.edu}








\renewcommand{\shortauthors}{A.K. Veldanda, et al.}


\fancyhead{}

\begin{abstract}
This paper proposes a novel two-stage defense ({\bf NNoculation}) against backdoored neural networks (BadNets) that, repairs a BadNet both pre-deployment and online in response to backdoored test inputs encountered in the field. In the pre-deployment stage, \sys retrains the BadNet with \emph{random} perturbations of clean validation inputs to partially reduce the adversarial impact of a backdoor. 
Post-deployment, \sys detects and quarantines backdoored test inputs by recording disagreements between the original and pre-deployment patched networks. A CycleGAN is then trained to learn transformations between clean validation and quarantined inputs; i.e., it learns to add triggers to clean validation images. Backdoored validation images along with their \emph{correct} labels are used to further retrain the pre-deployment patched network, yielding our final defense. 
Empirical evaluation on a comprehensive suite of backdoor attacks show that \sys outperforms all state-of-the-art defenses that make restrictive assumptions and only work on specific backdoor attacks, or fail on adaptive attacks. In contrast, \sys makes minimal assumptions and provides an effective defense, even under settings where existing defenses are ineffective due to attackers circumventing their restrictive assumptions.

\vspace{-1em}

\end{abstract}


\vspace{-1em}

\begin{teaserfigure}
    \includegraphics[width=\textwidth]{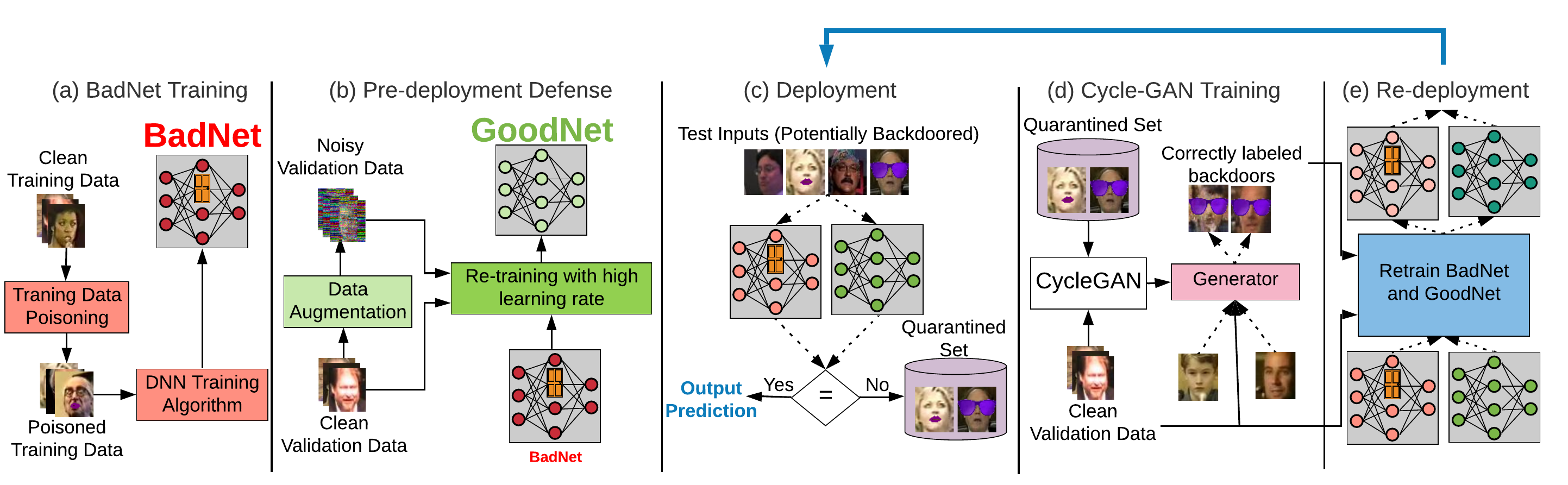}
    \vspace{-2.5em}
    \caption{An overview of \sys: (a) BadNet Training: First, the user/defender acquires a potential BadNet. (b) Pre-deployment Defense: The BadNet is retrained with noise-augmented validation data and high learning rate to obtain GoodNet. (c) Deployment: The BadNet and GoodNet are deployed as an ensemble that will either quarantine a test input if the outputs disagree or give a prediction if outputs agree. The quarantined set will mostly comprise poisoned inputs. (d) CycleGAN Training: A CycleGAN is trained on clean validation data and quarantined samples to output a generator that learns to generate correctly labeled poisoned inputs. (e) Re-deployment: Correctly labeled backdoor inputs are used to retrain the BadNet and GoodNet which are then deployed in the field, thus repeating the cycle from step (c). Dark red (BadNet) and dark green (GoodNet) denote networks with high and low attack success rates respectively.}
    \label{fig:overview}
\end{teaserfigure}

\maketitle

\section{Introduction}
\label{sec:intro} 

There is growing concern about the vulnerability of deep learning to backdooring (or Trojaning) attacks~\cite{badnets,neuraltrojans,sunglassesattack,poisonfrogs,cla_kang}, wherein an adversary compromises a deep neural network's (DNN) training data and/or training process with malicious intent. The vulnerability arises because users often do not have the computational resources to train complex models and/or the ability to acquire large, high-quality labeled training datasets required for high accuracy~\cite{dataarticle, skincancerdata, datasurvey}. Thus, users either outsource DNN training to untrusted third-party clouds or source {pre-trained} DNN models from online repositories like the Caffee Model Zoo~\cite{modelzoo, caffemodelzoo} or GitHub. This opens the door to DNN backdooring~\cite{badnets,neuraltrojans,sunglassesattack,poisonfrogs}: an adversary can train and upload a DNN model that is highly accurate on clean inputs (and thus on the user's validation set), but misbehaves when inputs contain special attacker-chosen backdoor triggers. Such maliciously trained DNNs have been referred to as ``BadNets." For example, Gu \textit{et al.}~\cite{badnets} demonstrated a traffic sign BadNet with state-of-the-art accuracy on regular inputs, but that classifies a stop sign with a Post-it note as a speed-limit sign.

\begin{table}[]
\tiny
\centering
\caption{Comparison of existing defenses in terms of the restrictive assumptions they make on trigger size and shape (TSS), existence of backdoor neurons (BN) that exclusively encode backdoors, and impact of the attack being limited to a single target label (A21).}
\vspace{-1em}
\label{tab:assumptions}
\resizebox{\columnwidth}{!}{%
\begin{tabular}{@{}lccccccc@{}}
\toprule
\multicolumn{1}{c}{\multirow{2}{*}{BadNet Defense}} & \multicolumn{3}{c}{Restrictive Assumptions} \\
\cmidrule(lr){2-4}
\multicolumn{1}{c}{} & \multicolumn{1}{c}{TSS} & \multicolumn{1}{l}{BN} & \multicolumn{1}{c}{A21} \\ \midrule
NeuralCleanse~\cite{neuralcleanse} & \rt{\CIRCLE}& \Circle & \rt{\CIRCLE} \\
Generative Modeling~\cite{duke} & \rt{\CIRCLE}& \Circle & \rt{\CIRCLE} \\
Fine Pruning~\cite{finepruning} & \Circle & \rt{\CIRCLE}& \Circle \\
ABS~\cite{absliu} & \Circle & \rt{\CIRCLE}& \rt{\CIRCLE} \\
STRIP~\cite{strip2019acsac} & \Circle & \Circle & \rt{\CIRCLE}\\
MNTD~\cite{mntd} & \rt{\CIRCLE} & \Circle & \Circle\\
\hline
\textbf{NNoculation} (This Work) & \Circle & \Circle & \Circle \\ \bottomrule
\end{tabular}%
}
\vspace{-2em}
\end{table}

Recent research has sought to mitigate the backdooring threat by detecting backdoored models or inputs (e.g.,~\cite{absliu,strip2019acsac}), and/or disabling backdoors (e.g.,~\cite{neuralcleanse,duke,finepruning}). However, existing defenses make restrictive assumptions (~\autoref{tab:assumptions}) that are easily circumvented by an adversary~\cite{veldanda2020evaluating}. One line of work~\cite{neuralcleanse,duke} makes strong assumptions about the size and shape of the trigger, i.e,  trigger is small, regularly shaped and super-imposed on the image~\cite{neuralcleanse} or that the trigger size and shape are \emph{known} to the defender~\cite{duke}. A second line of work~\cite{absliu, finepruning} assumes that the BadNet encodes the presence of a trigger using one (ABS~\cite{absliu}) or a few ~\cite{finepruning} dedicated neurons referred to as ``backdoor neurons." Finally, all defenses except Fine-pruning assume an ``all-to-one" attack, i.e., the BadNet mis-classifies \emph{any} backdoored input as belonging to a \emph{single} attacker chosen class. However, these defenses do not scale to a broader range of attacker objectives including ``all-to-all" attacks~\cite{badnets} wherein the target class depends on the class of the input. STRIP~\cite{strip2019acsac}, a recent defense, relies on this assumption. Because of these in-built assumptions, these prior works have restricted applicability and are susceptible to adaptive attacks, as we show in \autoref{subsec:prior_results}.

In this paper, we propose \sys, a new, general, defense against DNN backdooring attacks that relaxes the restrictive assumptions in prior work. Like NeuralCleanse and Fine-pruning, \sys also seeks to recover the backdoor trigger and re-train the BadNet with poisoned but \emph{correctly} labeled data, thus \emph{unlearning} bad behaviour. The challenge, however, is that the attacker has an asymmetric advantage, i.e., choosing from the vast space of backdoor patterns as long as they are not in the defender's validation data. Existing defenses mitigate this asymmetry by narrowing the search space of triggers via the restrictive assumptions listed in Table~\ref{tab:assumptions}. Our key observation is that the defender has a unique opportunity to level the playing field \emph{post-deployment}. That is, the test inputs to a deployed BadNet under attack must contain actual triggers; if the defender can identify even a fraction of backdoored test inputs, the search space of triggers can be narrowed considerably.  

Based on this observation, \sys patches BadNets in \emph{two} phases: 1) \emph{pre-deployment} using clean validation data (as in prior work), and 2) \emph{post-deployment} by monitoring test inputs, as depicted in~\autoref{fig:overview}. 
In the pre-deployment defense, \sys avoids assumptions about the trigger shape, size or location and instead retrains the BadNet with \emph{randomly perturbed} validation data using a high learning rate (see~\autoref{sec:pre-deploy}). That is, instead of defending against specific triggers, we seek robustness against a broad range of untargeted perturbations from clean inputs. 
Our pre-deployment defense 
yields a patched DNN, which we refer to as a "GoodNet," that reduces the attack success rate to between $\sim$0\% and $\sim$43\% on BadNets, even where \textbf{existing defenses are ineffective}.

Post-deployment, we use the GoodNet from the previous step to identify possible poisoned inputs, i.e., those on which the BadNet and GoodNet differ. These inputs are \emph{quarantined} and, over time, yield a dataset of inputs containing triggers (~\autoref{fig:overview}(c)). We then train a CycleGAN  to convert images from the domain of  clean validation data to that of the quarantined data. In other words, we teach the CycleGAN to add triggers to clean validation data  (~\autoref{fig:overview}(d)). Thus, we obtain a dataset of backdoored inputs with high-quality triggers and their corresponding clean labels (see detailed description in~\autoref{subsec:post_method}). We then re-train both the BadNet and GoodNet using this dataset, and redeploy the patched networks  (~\autoref{fig:overview}(e)). Our final defense reduces the attack success rate  down to $0\%-3\%$ with minimal loss in classification accuracy (see~\autoref{subsec:post_results}). 
Our post-deployment defense can be invoked multiple times in the field, 
thus enabling \sys to learn and adapt to new backdoor triggers continuously. Our code is available at: 
\url{https://github.com/akshajkumarv/NNoculation}.

\textbf{Contributions}
Our new contributions in this paper are: 
\begin{itemize}
    \item We propose \sys, a novel end-to-end defense against BadNets with minimal assumptions on the attack modalities including trigger size, shape and location, and impact.
    \item To the best of our knowledge, \sys is the \emph{first} BadNet defense that uses a combination of offline and online methods to continuously learn and adapt to actual backdoor behaviour observed in the field. \sys reliably reverse engineers BadNet triggers over a range of attacks.
    \item The first detailed side-by-side evaluation of state-of-the-art backdoor defenses against a wide range of backdoor attacks, ranging from a simple attack (single trigger, single target) to complex attacks that include multiple triggers and multiple targets. Comparisons of \sys with prior work show that it is the only defense that works comprehensively across a range of attacks, while prior defenses fail completely when their narrow assumptions are violated.
\end{itemize}
 
\vspace{-1em}
\section{Threat Model}
\label{sec:bckgnd}

We adopt the backdooring threat model from prior work~\cite{badnets, neuralcleanse, absliu, neuraltrojans, finepruning, duke}. We model two parties: a \textit{user} (or defender) who seeks to deploy a DNN for a target application by sourcing a \textit{trained} model for that application from an untrusted party, the \textit{attacker}. The attacker abuses the training process and returns a backdoored DNN, which misbehaves on images with backdoor triggers but otherwise provides acceptable performance on ``clean'' inputs. In this model, the user only has access to a small set of clean, correctly labeled images---primarily for validating the quality of the sourced model---that they can use to detect or remove backdoors. Note that the backdooring threat model is \emph{stronger} than that used in data poisoning attacks~\cite{data_poisoning1, data_poisoning2, data_poisoning3, data_poisoning4} addressed in literature. First, in data poisoning attacks, the user trains the model, while in backdooring the attacker controls the training process. Second, in data poisoning, the user has access to the training set, containing both clean and poisoned images (but doesn't know which is which)~\cite{certified_defense}, while in backdooring, the user only has a small validation set of clean images. 
Our assumptions about the attacker's and defender's goals and capabilities in the backdooring threat model are detailed below.

\vspace{-0.5em}
\subsection{Attacker's Goal and Capabilities} 

The \textit{attacker} has access to large and high-quality clean training dataset $\mathcal{D}_{tr}^{cl}$ drawn from distribution $\mathcal{P}^{cl}$. Let $f_{{\theta}_{cl}}$ denote the DNN obtained by benignly training on $\mathcal{D}_{tr}^{cl}$. The \textit{attacker} instead seeks to train a BadNet $f_{{\theta}_{bd}}$ that agrees with $f_{{\theta}_{cl}}$ on input $x$ drawn from $\mathcal{P}^{cl}$, \emph{but} misbehaves when $x$ is modified using a trigger insertion function $x^{p} =$ \texttt{poison}($x$). Misbehaviour in targeted attack can be represented as $\argmax f_{{\theta}_{bd}}(x^{p}) = \mathcal{T}(x^{p})$ where $\mathcal{T}(x^{p})$ is an attacker-chosen class which may be different from the benign DNN's prediction. For instance, in prior work, if $\mathcal{T}(x^{p}) = \mathcal{T}^{*}$, then all backdoor inputs are (mis)classified as a single target label $\mathcal{T}^{*}$. 

As in prior work~\cite{badnets, finepruning, neuralcleanse, absliu}, the attacker achieves this goal via training data poisoning. Specifically, the \textit{attacker} prepares $f_{{\theta}_{bd}}$ by training on both $\mathcal{D}_{tr}^{cl}$ and a set of poisoned inputs, $\mathcal{D}_{tr}^{bd\_p}$ which are prepared using the trigger insertion function \texttt{poison}($\cdot$). 
The hyper-parameters of the attack include the fraction, $p$, of poisoned training data, in addition to the hyper-parameters of the training process. The attacker optimizes the attack hyper-parameters such that $f_{\theta_{bd}}$ maintains good accuracy on $\mathcal{D}_{valid}^{cl}$ but reliably misbehaves on $\mathcal{D}_{valid}^{p}$. Once an unsuspecting \textit{user} deploys the BadNet, the \textit{attacker} triggers the DNN misbehavior by providing poisoned test data $x^{p}$ containing the backdoor trigger.

\vspace{-0.5em}
\subsection{Defender's Goals and Capabilities}

The \textit{user} (referred to interchangeably as the \textit{defender}) wishes to deploy a DNN for the application advertised by the \textit{attacker}, but does not have the resources to acquire a large, high-quality dataset for it. Instead, the \textit{user} downloads the DNN, $f_{{\theta}_{bd}}$, uploaded by the \textit{attacker}, and uses a small validation dataset, $\mathcal{D}_{valid}^{cl}$, of clean inputs to verify the DNN's accuracy. In addition, the user seeks to patch $f_{{\theta}_{bd}}$ to eliminate backdoors---ideally, the patched DNN should output correct labels for backdoored inputs, or detect and classify them with a warning. 

To meet these goals, the \textit{user} has access to two assets pre-deployment: full, white-box access to $f_{{\theta}_{bd}}$ and a small clean validation dataset $\mathcal{D}_{valid}^{cl}$. Post-deployment, the \textit{user} also has full access to the online stream of inputs $\mathcal{D}_{stream}$ seen by the deployed model. As in prior work~\cite{neuralcleanse,absliu}, we do not bound (but will seek to minimize) the \textit{user}'s computational effort, i.e., the \textit{user}'s primary limitation is the paucity of high-quality training data, not computational resources. 

\vspace{-0.5em}
\subsection{Security Metrics}

We evaluate our defense successes based on the following two metrics, evaluated using a held-out test dataset $\mathcal{D}_{test}$ that emulate post-deployment inputs: (1) Clean Data Accuracy (CA), defined as the percentage of clean test data that is correctly classified; (2) Attack Success Rate (ASR), percentage of backdoored test images classified as backdoors. Our defense seeks to lower ASR (reducing power held by \textit{attacker}) while minimizing impact on CA.

\section{\sys Defense}
\label{sec:method}

In this section, we describe \sys in more detail.

\subsection{Overview}

We begin with a high-level overview of \sys. \sys is a two stage defense. First, the user (defender) acquires a DNN---a potential BadNet $f_{{\theta}_{bd}}$. In the first stage, i.e., the \textit{pre-deployment stage}, the defender retrains $f_{{\theta}_{bd}}$ with an augmented dataset containing both clean validation data $\mathcal{D}_{valid}^{cl}$ \textit{and} noisy versions of the clean input at a ``high" learning rate. Retraining with augmented data aims to stimulate a wide range of behaviours in the DNN, and forces the DNN to pay more attention to the unmodified portions of the image, thus reinforcing ``good" behaviours. In addition, a high learning rate nudges the BadNet away from its learned ``bad" behaviours. The result is a new DNN, $\theta_{gd}$, with reduced, but not zero, ASR. We then deploy $f_{{\theta}_{bd}}$ and $f_{{\theta}_{gd}}$ as an \emph{ensemble}. 
    
In the \textit{post-deployment stage}, inputs (in $\mathcal{D}_{stream}$) that disagree between $f_{{\theta}_{bd}}$ and $f_{{\theta}_{gd}}$ are marked as suspects and quarantined. As long as the pre-deployment reduces ASR (even if not down to zero), the quarantined dataset likely includes attacker-poisoned data. Now, using the clean validation dataset and quarantined dataset, we \emph{learn} the function \texttt{poison}($x$) using a CycleGAN $G$ that transfers between the two domains (in effect, the CycleGAN learns to poison clean data!). We then use the reverse-engineered trigger to retrain $\theta_{bd}$ and $\theta_{gd}$, this time with \emph{correctly} labeled examples of backdoors. 
This reduces ASR down to near zero, while preserving CA. Online retraining can be redeployed if further backdoor inputs are detected.

\subsection{Pre-Deployment Defense}
\label{sec:pre-deploy}

\begin{figure}[t]
    \centerline{\includegraphics[width=0.7\columnwidth]{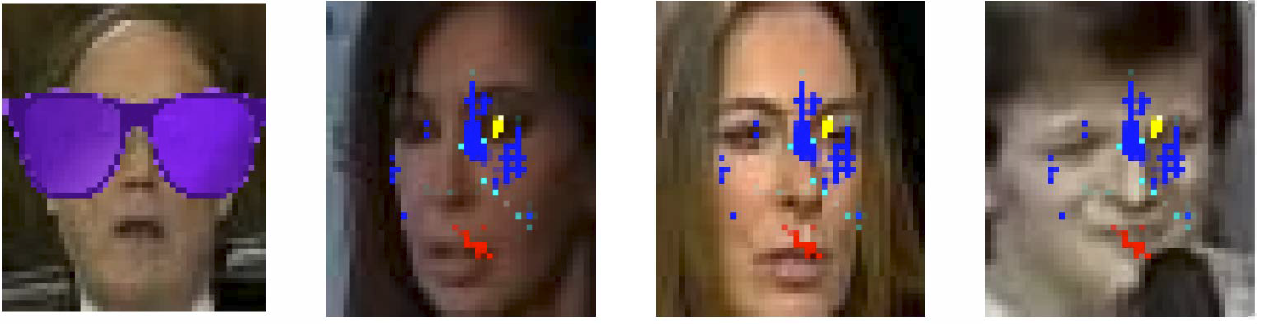}}
    \vspace{-1em}
    \caption{Shortcomings of NeuralCleanse: The leftmost image is the actual trigger; the other images are incorrectly reverse-engineered triggers by NeuralCleanse.}
    \label{fig:neural-fail}
    \vspace{-2em}
\end{figure}

\paragraph{Shortcomings of existing defenses:} Pre-deployment defense is motivated by shortcomings observed in prior work, specifically, Fine-pruning and NeuralCleanse. {Fine-pruning}~\cite{finepruning} is based on the observation that backdoor and clean inputs excite different neurons in a BadNet. The defense first prunes neurons unactivated by clean validation data, suspecting them of encoding backdoors, and then uses the clean validation data to re-train the pruned network. In our experiments, however, we found several BadNets wherein clean and backdoored inputs excited the \emph{same} neurons --- pruning these neurons resulted in large drops in clean CA that could not be recovered via retraining. Hence, \sys's pre-deployment defense eschews the initial pruning step, but instead, performs re-training with noise augmentation using a high learning rate.

\paragraph{Noise augmentation:}  Noise augmentation is motivated by shortcomings of NeuralCleanse~\cite{neuralcleanse}. This method seeks to directly recover triggers from BadNets by finding the smallest contiguous overlay that causes any input to be mis-classified as a unique target label. In practice, triggers need not be small or contiguous. For example, in~\autoref{fig:neural-fail} we illustrate the output from NeuralCleanse given BadNets triggered by a large, but semantically meaningful, sunglasses trigger for face recognition~\cite{finepruning}. The recovered triggers bear little resemblance to the original, missing its size, shape, and color. However, we find that in some instances, given oracular knowledge of the attacker's target label, NeuralCleanse is able to find small patches of the trigger, and that retraining with only a small part of the trigger reduces ASR, although not down to zero. 

Instead of reverse engineering (parts of) the trigger (which is time-consuming and rarely works~\cite{neuralcleanse}), in \sys we seek to cast a wide net by adding random noise to validation images, with the hope of catching some aspects of the trigger pre-deployment. Specifically, we use a noise augmentation function $A(\mathcal{D},\eta, \gamma)$ that randomly samples $\gamma$ fraction of pixels from an image and replaces them with values sampled from distribution $\eta$. We then add noise to images in the clean validation set, $\mathcal{D}_{valid}^{cl}$, using $A$ and re-train on clean and noise augmented validation sets.
    
\begin{table}[ht]
    \centering
    \vspace{-0.5em}
    \caption{Example of grid search algorithm for BadNet-FSA to pick $\theta_{gd}$. The baseline CA is 91.34\% and all networks with up to $\theta=3\%$ drop in CA are shown in bold font. The network picked is starred.}
    \vspace{-1em}
    \resizebox{\columnwidth}{!}{%
    \begin{tabular}{@{}cccccc@{}}
    \toprule
     $\gamma$ & $\alpha$=0.001 & $\alpha$=0.003 & $\alpha$=0.004 & $\alpha$=0.005 & $\alpha$=0.006 \\ 
    \midrule
    10\% & \textbf{92.55} & 86.79 & 85.3 & 81.86 & 83.9  \\
    20\% & \textbf{93.3} & \textbf{88.27} & 87.53 & 80 & 81.48 \\
    30\% & \textbf{93.39} & \textbf{88.46} & 85.3 & 87.72 & 83.72 \\
    40\% & \textbf{92.09} & \textbf{90.41} & 86.04 & 81.39 &  82.41\\
    50\% & \textbf{92.46}  & \textbf{89.02} & 87.06 & 82.13 &  83.62 \\
    60\% & \textbf{93.58} & \textbf{88.46} & \textbf{88.46*} & 86.13 & 84.65 \\ \bottomrule
    \end{tabular}
    }
    \label{tab:pre-fsa}
    \vspace{-2em}
\end{table}
    
\paragraph{Finding the optimum learning rate and noise level:} To coax the BadNet to unlearn its misbehaviour, we seek to retrain the BadNet with the \emph{highest possible learning rate and noise level} while keeping the drop in clean accuracy below a user-specified threshold $\theta$. To do so, we perform a \emph{grid search} over varying noise levels and learning rates, and (i) first select networks that  have clean accuracy within threshold $\theta$ of the BadNet's accuracy, then (ii) of these, we further pick networks trained with the largest learning rate, and finally (iii) of the remaining, we select the network with the highest noise level. An example of a grid search for BadNet-FSA is shown in~\autoref{tab:pre-fsa}. In this example, we found networks at $\alpha= {0.001, 0.003, 0.004} $ above our CA threshold, and picked the network with  $\alpha=0.004$ and $\gamma = 60\%$ as our pre-deployment GoodNet.

\subsection{Post-Deployment Defense}
\label{subsec:post_method}

The input to \sys's post-deployment defense is the patched network, $\theta_{gd}$, from the pre-deployment stage. In the field, we deploy $\theta_{gd}$ \emph{in parallel with} $\theta_{bd}$ to detect backdoored inputs in $\mathcal{D}_{stream}$ ---if the two disagree, we predict that the input is backdoored and output $\theta_{gd}$'s prediction, else, we output their common prediction. We refer to this parallel combination as an \emph{ensemble}.  

After deploying the \textit{ensemble}, the system begins to receive unlabeled data for classification, $\mathcal{D}_{stream}$. We assume that the attacker will try to attack the system---some fraction of $\mathcal{D}_{stream}$ includes poisoned data containing the trigger. The exact proportion of poisoned data is unknown to the defender. Since the clean accuracies of $f_{\theta_{bd}}$ and $f_{\theta_{gd}}$ are close, disagreements will arise largely from poisoned inputs, and thus we expect the quarantined dataset to have a reasonable fraction of backdoors. 

Once sufficiently many images, $\mathcal{N}$, have been collected in the quarantined dataset the defender trains a CycleGAN where domain 1 is represented by $\mathcal{D}_{valid}^{cl}$ and domain 2 by $\mathcal{D}_{quarantine}$. We use the CycleGAN architecture proposed by~\cite{CycleGAN2017} with nine residual blocks in its generator, $G$, and  a discriminator based on a 70×70 PatchGAN~\cite{patchgan}. The CycleGAN's complete architectural parameters are described in Appendix~\ref{app-sec:cyclegan_details_params}.

The resulting generator, $G$ approximates the attacker's \texttt{poison}($x$) function. Using $G$, the defender creates a new ``treatment" dataset by executing the CycleGAN's generator on clean validation inputs, $\mathcal{D}_{treat}=G(\mathcal{D}_{valid}^{cl})$, and labeling the resulting images with labels from the validation set.

Finally, the defender fine-tunes $\theta_{bd}$ and $\theta_{gd}$, using $\mathcal{D}_{treat}$ and $\mathcal{D}_{valid}^{cl}$ producing repaired networks, $\theta_{bd'}$ and $\theta_{gd'}$, respectively. These networks are then deployed, again as an ensemble, to enable continuous repair in response to new attacks. For example, consider a setting where the BadNet has multiple triggers, but the attacker only uses a subset of triggers in the initial attack and uses the remaining triggers after the first round of repair. Poisoned data with the new triggers will be quarantined by the repaired networks which will be treated again once sufficiently many instances of the new trigger are collected.

\begin{table*}[t]
\centering
\caption{Baseline BadNet Preparation: dataset split, trigger, and target label}
\vspace{-1em}
\resizebox{\textwidth}{!}{%
\begin{tabular}{cccC{2cm}C{2cm}C{2cm}C{2cm}C{2cm}cc}
\toprule
BadNet & Attack Setting & Dataset & \# of classes & Train samples per class & Valid samples per class & Online Stream + Test samples per class & Trigger & Target label \\  \midrule
AAA & All-All & MNIST & 10 & 5500 & 450 & 1000 & Patterned trigger & \textit{y} $\rightarrow$ \textit{y+1} \\ \midrule
TCA & Trigger Comb. & CIFAR-10 & 10 & 5000 & 500 & 450 & Yellow triangle $+$ red square & 7 \\ \midrule
PN & Simple & \multirow{2}{*}{GTSRB} & \multirow{2}{*}{43} & \multirow{2}{*}{820} & \multirow{2}{*}{90} & \multirow{2}{*}{270} & Post-it note & 0 \\
FSA & Feature Space &  &  &  &  &  & Gotham filter & 35 \\ \midrule
SG & Simple & \multirow{4}{*}{YouTube Face} & \multirow{4}{*}{1283} & \multirow{4}{*}{81} & \multirow{4}{*}{9} & \multirow{4}{*}{8} & sg & 0 \\ 
LS & Simple &  &  &  &  &  & ls & 0 \\
MTSTA* & Multi-trigger, single taget &  &  &  &  &  & ls, eb, sg & 4,4,4 \\
MTMTA* & Multi-trigger, multi-target &  &   &  &  &  & ls, eb, sg & 1,5,8 \\ \midrule
IN & Simple & ImageNet & 1000 & 1200 & 25 & 20 & Red square & 0 \\
\bottomrule
\end{tabular}%
}
\scriptsize{ \begin{flushleft} * for MTSTA and MTMTA, ls, eb, sg corresponds to lipstick, eyebrow, sunglasses trigger, respectively \end{flushleft}}
\label{tab:baseline-badnet}
\end{table*}

\section{Experimental Setup}
\label{sec:exps}

In this section, we describe our experimental setup. All our experiments are on a desktop computer with Intel CPU i9-7920X (12 cores, 2.90 GHz) and single Nvidia GeForce GTX 1080 Ti GPU.

\vspace{-1em}
\subsection{BadNet Preparation and Attack Settings}
\label{subsec:badprep}

\begin{figure*}[t]
    \centering
    \includegraphics[width=\textwidth]{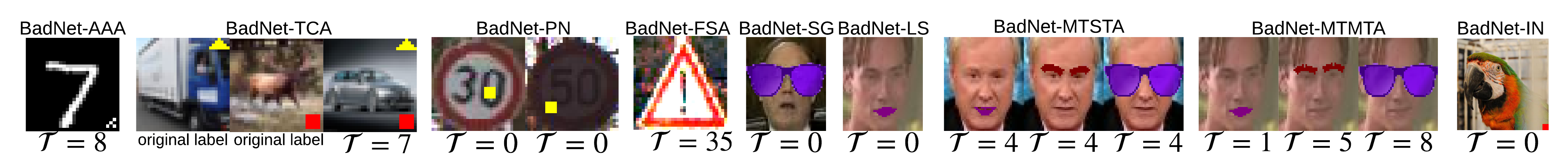}
    \vspace{-2em}
    \caption{Examples of the datasets, triggers, and target labels used in this study.}
    \label{fig:triggers}
    \vspace{-2em}
\end{figure*}

To comprehensively evaluate \sys, we prepare several BadNets on MNIST~\cite{mnistdataset}, CIFAR-10~\cite{krizhevsky2009learning}, German Traffic Sign Recognition Benchmark (GTSRB)~\cite{gtsrbdataset}, YouTube Aligned Face~\cite{youtubedataset}, and ImageNet datasets~\cite{imagenet_cvpr09}. We report the number of classes, and the number of training, validation and test samples per class for each dataset in ~\autoref{tab:baseline-badnet}. We implemented several previously reported and three new attacks (MTSTA, MTMTA and TCA) on these datasets using the attack training hyper-parameters reported in~\autoref{tab:base-hyper}. The attacks are described below and examples of triggers corresponding to each attack are shown in~\autoref{fig:triggers}. The baseline CA and ASR for the BadNets are reported in~\autoref{tab:comparision_results}.

\textbf{\textbf{All-to-One Attack}}: Any poisoned input will be classified as the attacker chosen target label. BadNet-SG and BadNet-LS are trained on YouTube Aligned Face dataset using sunglasses (sg) and lipstick (ls) as the trigger respectively. Similarly, BadNet-PN is trained on GTSRB with variable location post-it note trigger. Additionally, to evaluate the scalability of \sys on large datasets and complex architecture, we train BadNet-IN with a DenseNet-121~\cite{densenet} architecture using ImageNet dataset, choosing a
red square as the trigger. All these BadNets have target label, $\mathcal{T} = 0$, consistent with prior work (e.g., as in~\cite{finepruning,absliu}).

\textbf{\textbf{All-All Attack (AAA)}~\cite{badnets}}: In the presence of the trigger, the backdoor will cause an input with ground truth label \textit{y} to be classified with target label \textit{y+1}. BadNet-AAA is trained on MNIST dataset using a fixed pixel pattern trigger.

\textbf{\textbf{Multi-Trigger Single-Target Attack (MTSTA)}}: The attacker uses multiple triggers but the backdoor has only one target label. BadNet-MTSTA is trained on YouTube Aligned Face dataset using three triggers: lipstick (ls), eyebrow highlighter (eb) and sunglasses (sg). All these triggers will activate the backdoor to classify a poisoned input with the target label, $\mathcal{T} = 4$.

\textbf{\textbf{Multi-Trigger Multi-Target Attack (MTMTA)}}: The attacker uses multiple triggers, where poisoning any input with one of the triggers will cause the BadNet to classify the input with the trigger's corresponding target. BadNet-MTMTA is trained on YouTube Aligned Face dataset using three triggers with three corresponding target labels: ls ($\mathcal{T} = 1$), eb ($\mathcal{T} = 5$) and sg ($\mathcal{T} = 8$).

\textbf{\textbf{Feature Space Attack (FSA)~\cite{absliu}}}: The attacker uses transformations in input space that lead to patterns in feature space. BadNet-FSA is trained on GTSRB using Gotham Filter as the feature space trigger. Applying the filter to any input will cause the BadNet to output the target label, $\mathcal{T} = 35$.

\textbf{\textbf{Trigger Combination Attack (TCA)}}: In this attack, the backdoor is activated only in the presence of a \textit{combination} of triggers; it ignores any trigger appearing by itself. BadNet-TCA is trained on CIFAR-10 dataset using a combination of red square and yellow triangle as the trigger and $\mathcal{T} = 7$.

\begin{table}[ht]
\centering
\caption{Training hyper-parameters of baseline BadNets \label{tab:base-hyper}}
\vspace{-1em}
\resizebox{\columnwidth}{!}{%
\begin{tabular}{@{}lccccccc@{}}
\toprule
 & \multirow{2}{*}{AAA} &  \multirow{2}{*}{CLA} & \multirow{2}{*}{TCA} & \multirow{2}{*}{PN, FSA} & SG, MTSTA, & \multirow{2}{*}{IN} \\
 &  &  &  &  & LS, MTMTA &  \\
 \midrule
Architecture & ~\cite{strip2019acsac} & Custom & DeepID~\cite{deepid} & ~\cite{neuralcleanse} & NiN~\cite{networkinnetwork} & DenseNet-121~\cite{densenet} \\
batch size & 32 & 32 & 128 & 32 & 1283 & 256\\
epochs & 50 & 25 & 200 & 15 & 200 & 12\\
learning rate & 1e-4 & 1 & 0.01* & 1e-3 & 1 & 0.01**\\
optimizer & Adam & Adadelta & SGD & Adam & Adadelta & SGD\\
preprocessing & 1./255 & 1./255 & $\frac{(\cdot) - \mu}{\sigma}$ & 1./255 & 1./255 & $\frac{(\cdot) - \mu}{\sigma}$\\ \bottomrule
\end{tabular}
}
\scriptsize{\begin{flushleft} * scheduler: lr = 0.01 if epoch $\leq 80$; 0.005 if $80<$ epoch $ \leq 140$; else 0.001 \\ ** scheduler: lr = 0.01 if epoch $\leq 4$; 0.001 if $4<$ epoch $ \leq 8$; else 0.0001\end{flushleft}} 
\vspace{-2em}
\end{table}

\vspace{-0.75em}
\subsection{\sys Parameters}
\subsubsection{Set-up of Pre-deployment Defense}

We use the Python \texttt{imgaug}~\cite{imgaug} library to prepare our noise augmented datatsets using a Gaussian noise distribution, $\eta$, ($mean=128$ and $variance=51.2$) which are the default out-of-the-box options in the library. The noise fraction $\gamma$ varies from $10\%$-$60\%$ in increments of $10\%$. The initial learning rate for pre-deployment training, $\alpha_{0}$, is set to the original learning rate of the corresponding BadNet and increased in multiples thereafter.

\vspace{-0.5em}

\subsubsection{Set-up for Post-deployment Defense}
Post-deployment defense is triggered after the first $\mathcal{N} = 200$ quarantined images are collected, at which point the CycleGAN~\cite{CycleGAN2017} is trained on the quarantined dataset collected thus far and 500 images from $\mathcal{D}_{valid}^{cl}$. All networks in the CycleGAN are trained from scratch for 200 epochs using Adam optimizer~\cite{adam_optim}, with an initial learning rate of 0.0002 with linear decay after the first 100 epochs~\cite{CycleGAN2017, cyclegan_implementation}. More details about CycleGAN network architecture and training details are presented in Appendix~\ref{app-sec:cyclegan_details_params}.

\subsection{Baselines for Comparison}

We compare \sys with NeuralCleanse~\cite{neuralcleanse}, Fine-pruning~\cite{finepruning}, STRIP~\cite{strip2019acsac}, and Qiao et al.'s approach~\cite{duke} using their reference implementations where possible. NeuralCleanse and Qiao et al. attempts to identify the attacker's target label. However, NeuralCleanse identifies the \emph{correct} target label for only 2 out of 8 BadNets and Qiao et al. fails on all BadNets. In the situations where NeuralCleanse and Qiao et al.'s defenses are unable to determine the target label, we endow those defenses with ``oracular knowledge" of the target label so that they can generate a result. We used FAR/FRR for STRIP to be consistent with the paper, but these can also be easily translated to ASR and CA ($ASR_{STRIP} = ASR_{Baseline} \times FAR$ and $CA_{STRIP} = CA_{Baseline} \times (100 - FRR)$). Finally, two recent defenses, ABS~\cite{absliu} and the very recent Meta Neural Trojan Detection (MNTD)~\cite{mntd}, are different from previous defenses (including \sys) because they detect whether an entire network is backdoored. We compare them separately. Also, ABS provides an executable that works on CIFAR-10, and thus we could compare \sys with ABS only on BadNet-TCA.

\section{Experimental Results}
\label{sec:results}

\begin{table*}[t]
\centering
\caption{Performance of NNoculation (with 3\% Threshold) on baseline BadNets in comparison with prior work. Final \sys solution corresponds to the post-deployment defense and the results highlighted in bold font correspond to Pareto optimal solutions}. 

\label{tab:comparision_results}
\vspace{-1em}
\resizebox{\textwidth}{!}{%
\begin{tabular}{cccccccccccccc}
\toprule
\multicolumn{1}{l}{} & \multicolumn{2}{c}{BadNet (Baseline)} & \multicolumn{2}{c}{\sys (pre-)} & \multicolumn{2}{c}{\sys (post-)} & \multicolumn{2}{c}{Fine-Pruning} & \multicolumn{2}{c}{NeuralCleanse} & STRIP (FRR=5\%) & \multicolumn{2}{c}{Qiao \textit{et al.}~\cite{duke}} \\
BadNet & CA & ASR & CA & ASR & CA & ASR & CA & ASR & CA & ASR & FAR & CA & ASR \\
\cmidrule(lr){1-1} \cmidrule(lr){2-3} \cmidrule(lr){4-5} \cmidrule(lr){6-7} \cmidrule(lr){8-9} \cmidrule(lr){10-11} \cmidrule(lr){12-12} \cmidrule(lr){13-14}
MNIST-AAA & 97.76 & 95.91 & 95.3 & 0 & \textbf{96.21} & \textbf{0} & 91.31 & 2.17 & \multicolumn{2}{c}{Fails} & 98.39 & \multicolumn{2}{c}{Fails} \\
\midrule
CIFAR10-TCA & 87.71 & 99.9 & 83.6 & 4.36 & \textbf{84.14} & \textbf{2.31} & 78.77 & 42.28 & 88.59$^\dag$ & 99.82$^\dag$ & 22.22 & \multicolumn{2}{c}{out of scope} \\
\midrule
GTSRB-PN & 95.46 & 99.82 & 92.9 & 13.37 & \textbf{93.01} & \textbf{0} & 92.25 & 24.12 & \textbf{95.24} & \textbf{12.39} & 90.28 & \multicolumn{2}{c}{out of scope} \\
GTSRB-FSA & 95.08 & 90.06 & 93.26 & 3.28 & \textbf{92.3} & \textbf{3.24} & 88.38 & 3.79 & \textbf{95.8} & \textbf{28.99} & 98.34 & 93.65$^\dag$ & 5.58$^\dag$ \\
\midrule
YouTube-SG & 97.89 & 99.98 & 94.18 & 35.22 & \textbf{92.03} & \textbf{0} & 91.40 & 30.12 & 95.74$^\dag$ & 38.09$^\dag$ & 7.6 & 77.64$^\dag$ & 1.88$^\dag$ \\
YouTube-LS & 97.19 & 91.51 & 93.99 & 32.34 & \textbf{93.15} & \textbf{0} & 91.58 & 1.67 & 97.14$^\dag$ & 28.44$^\dag$ & 16.11 & \multicolumn{2}{c}{out of scope} \\
YouTube-MTSTA* & 95.84 & \{92.2, 92.2, 100\} & 92.81 & \{36.5, 3.1, 0\} & \textbf{92.37} & \{1.86, \textbf{0, 0}\} & \textbf{91.61} & \{\textbf{0}, 2, 68.3\} & 93.37$^\dag$ & \{0, 0, 8.7\}$^\dag$ & \{11, 53.7, 4.9\} & \multicolumn{2}{c}{out of scope} \\
YouTube-MTMTA* & 95.93 & \{91.5, 91.4, 100\} & 92.2 & \{30, 7.1, 13.6\} & \textbf{91.48} & \{\textbf{0, 0}, 1.2\} & \textbf{90.36} & \{2.2, 13.8, \textbf{0}\} & 94.18$^\dag$ & \{30.8, 0, 95.7\}$^\dag$ & \{14.1, 44.3, 3\} & \multicolumn{2}{c}{out of scope} \\
\bottomrule
\end{tabular}
}
\scriptsize{ \begin{flushleft} * for MTSTA and MTMTA, the ASR corresponds to using lipstick, eyebrow, sunglasses trigger, respectively. $^\dag$we give oracular knowledge to these defenses \end{flushleft}}
\vspace{-1em}
\end{table*}

\begin{figure*}[ht]
\centering
\subfloat[BadNet-SG CA (pre-)]{\label{fig:sg-ca}\includegraphics[width=0.33\textwidth]{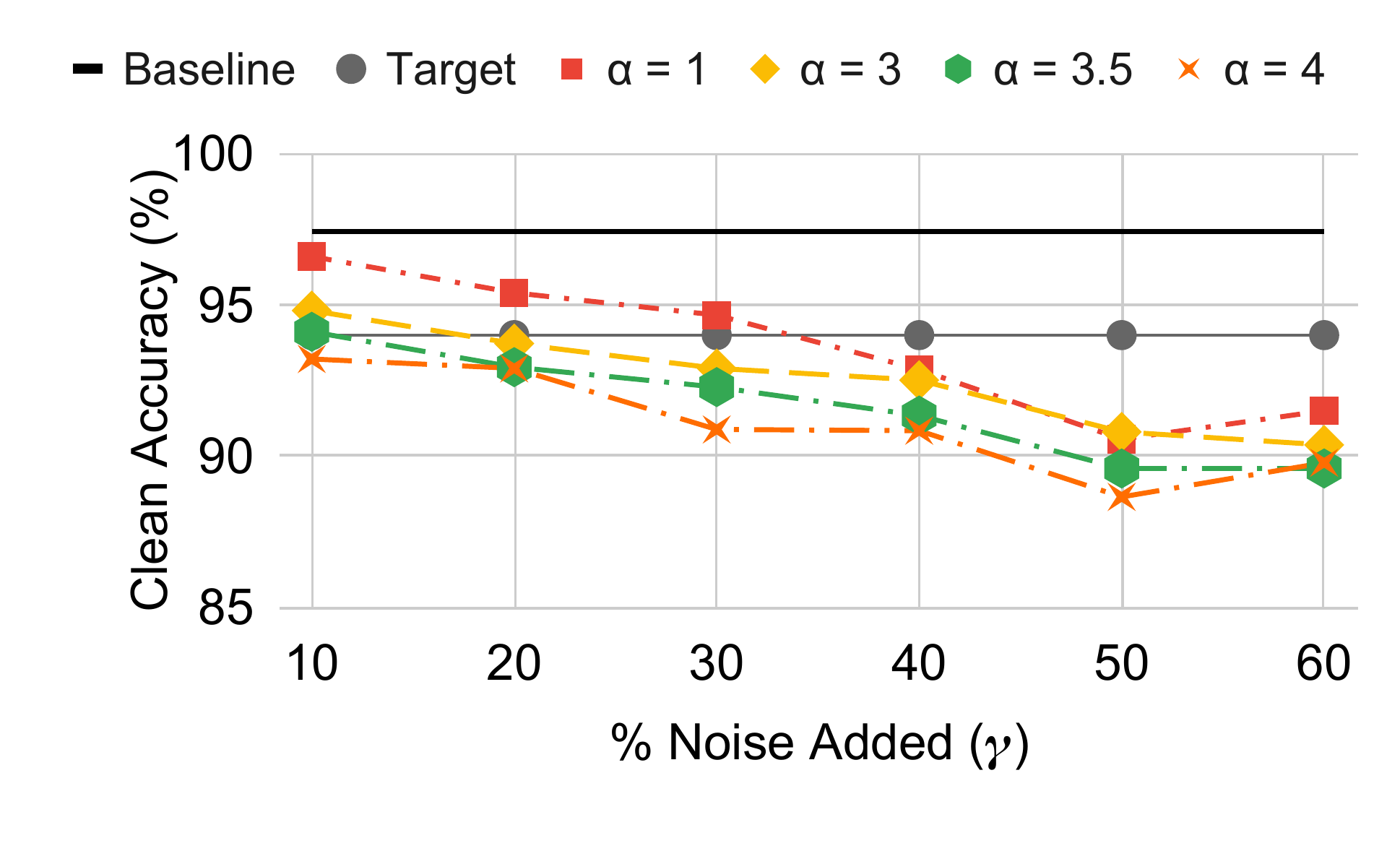}}
\subfloat[BadNet-SG ASR (pre-)]{\label{fig:sg-asr}\includegraphics[width=0.33\textwidth]{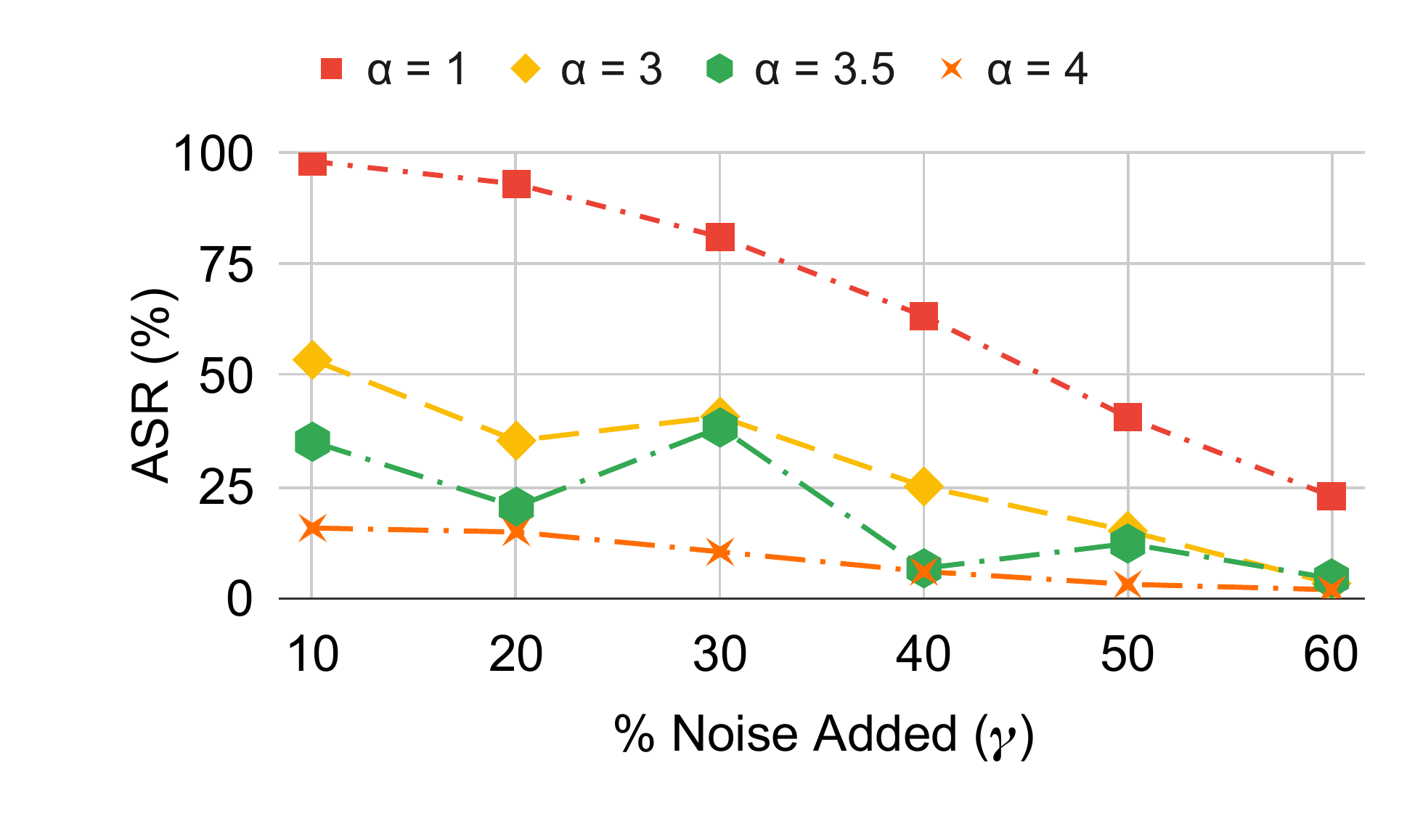}}
\subfloat[BadNet-SG CA and ASR (post-)]{\label{fig:sg-pcd}\includegraphics[width=0.33\textwidth]{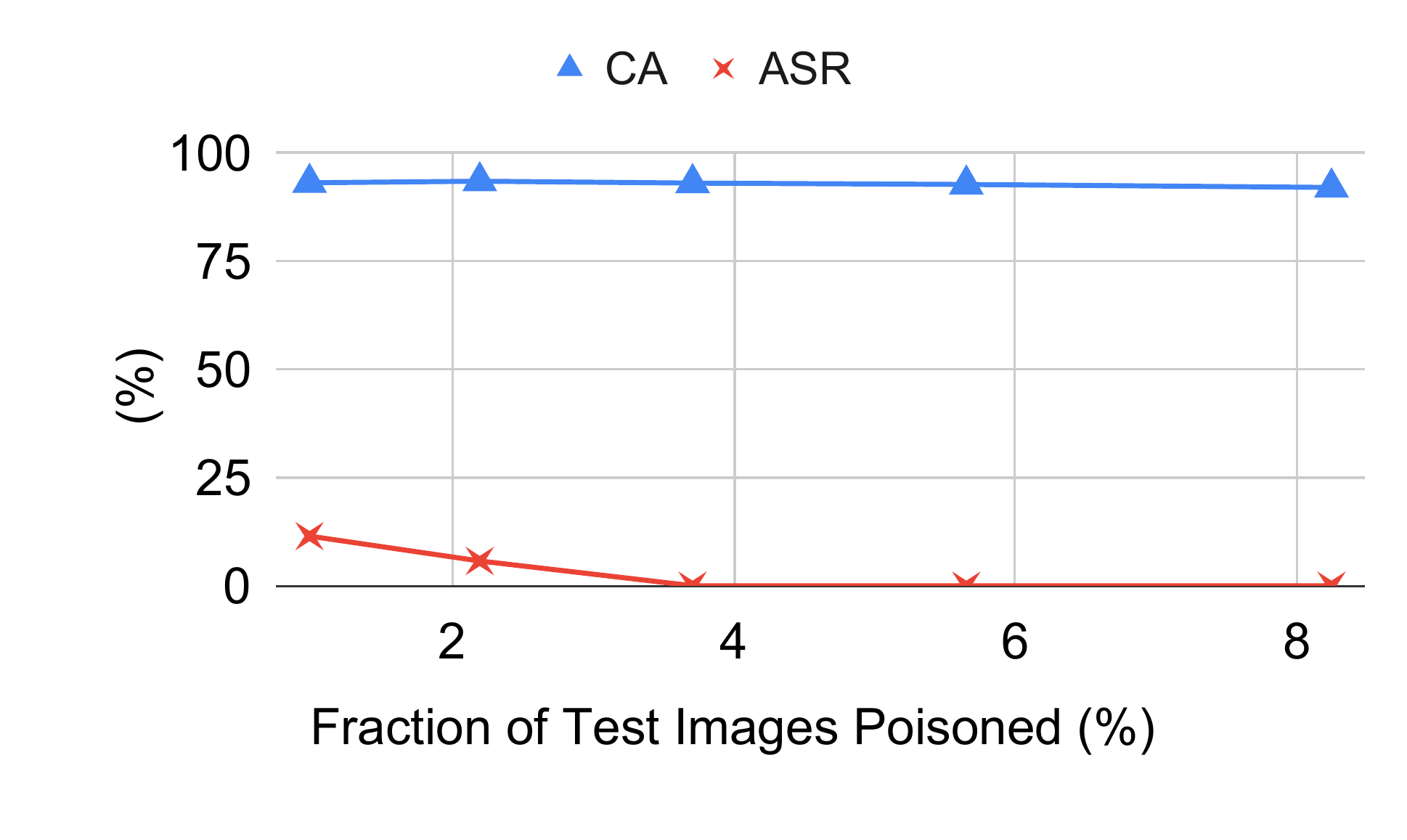}}
\vspace{-1em}
\caption{(a) and (b) show the effect of pre-deployment treatment on CA (on evaluation data) and ASR (on test data), respectively, under varying learning rate ($\alpha$) and noise ($\gamma$) settings for BadNet-SG. (c) Effect of post-deployment treatment on CA and ASR (both measured on test data) on BadNet-SG from re-training with data produced by the CycleGAN prepared with quarantined data for varying \textit{poison/clean input data stream ratios} within $\mathcal{D}_{stream}$.
\label{fig:pre-deployment-ca-asr}}
\vspace{-1em}
\end{figure*}

\subsection{Efficacy of Pre-deployment Defense}

\autoref{fig:pre-deployment-ca-asr} plots the CA and ASR of our pre-deployment defense on BadNet-SG for varying learning rates ($\alpha$) and noise levels ($\gamma$).~\autoref{tab:comparision_results} presents additional results of pre-deployment treatment on all remaining BadNets. The results are qualitatively similar (see Appendix~\autoref{fig:appendix_pre-deployment-ca-asr}).

Across all experiments, increasing $\alpha$ and $\gamma$ results in a drop in CA (ranging from 0.16\% to the largest drop of 15.63\%) and a reduction in ASR (in some case down to 0\%).  Varying $\alpha$ and $\gamma$ allows one to balance ASR reduction and CA loss. For all BadNets, there is at least one parameter setting that provides an ASR (below $44\%$) and CA ($\sim3\%$ less than baseline) that makes our pre-deployment defense alone competitive with prior works. NNoculation’s pre-deployment defense is Pareto optimal for 7/8 BadNets (except BadNet-PN), whereas Fine-Pruning, Neural Cleanse (without oracle access) and STRIP are pareto optimal for 3/8 , 2/8 and 1/8 respectively. More importantly, the pre-deployment defense results in a large enough drop in ASR to ensure the success of the post-deployment step. 

\subsection{Efficacy of Post-deployment Defense}
\label{subsec:post_results}

\begin{figure}[ht]
    \centering
    \includegraphics[width=\columnwidth]{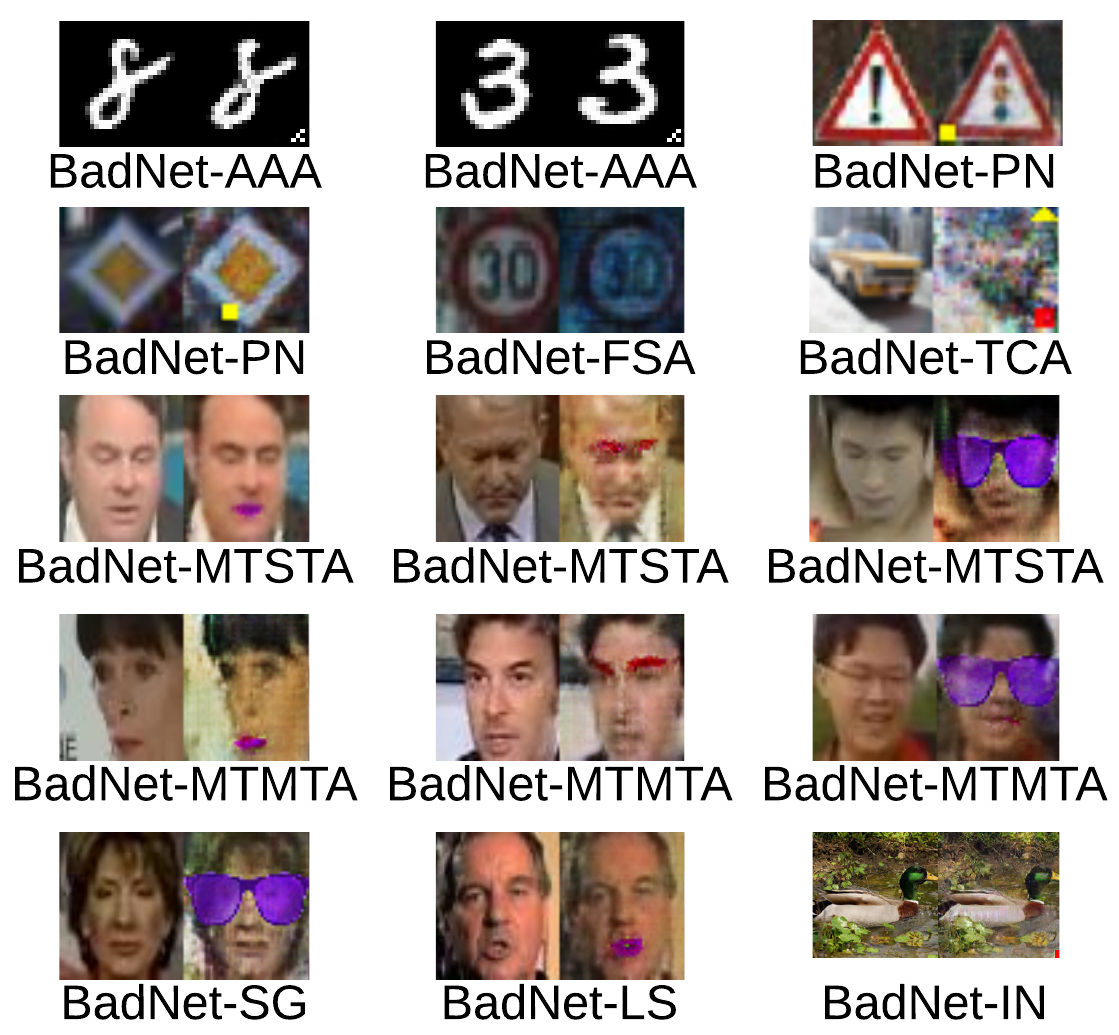}
    \vspace{-2em}
    \caption{Examples of the CycleGAN-based trigger reverse-engineering. For each BadNet, the left image corresponds to the clean input and the right image corresponds to the poisoned image generated by the CycleGAN.\label{fig:cyclegan_rev_triggers}}
    \vspace{-2em}
\end{figure}

To understand \sys's post-deployment defense, we show examples of the reverse-engineered triggers produced by the CycleGAN in~\autoref{fig:cyclegan_rev_triggers}. Across the board, \sys faithfully learns to add triggers to clean images, even though it has access to only 200 quarantined images. This is consistent with results from the original CycleGAN paper~\cite{CycleGAN2017} that achieved high-quality reconstructions on small datasets with as few as 400 images, for example, on the facades $\leftrightarrow$ photographs and artistic style transfer tasks, especially where domains are similar. 

Although we can obtain higher quality reconstructions with larger quarantined datasets (see~\ref{app-sec:cyclegan_quality_training_data}), these are not needed because exact trigger reconstruction is not necessary for backdoor unlearning~\cite{neuralcleanse, absliu}. Thus we are able to keep the size of the quarantined dataset small.
    
\sys works even if the attacker poisons a relatively small fraction of $\mathcal{D}_{stream}$. For BadNet-SG, we observe that as the attacker poisons more than $4\%$ of inputs in $\mathcal{D}_{stream}$, \sys's ASR drops to almost zero (see ~\autoref{fig:sg-pcd}), while the clean accuracy remains relatively constant. In other words, \sys forces the attacker into an unfavorable trade-off: either poison a very small fraction of inputs, or poison a larger fraction of inputs which are then detected by \sys with close to $100\%$ accuracy. That is, the attacker's \emph{effective} attack success rate, the ASR times the fraction of poisoned test inputs, is small across the board. In the next section, we evaluate \sys on all our BadNets assuming that $20\%$ of test inputs are poisoned.

\subsection{Comparisons with prior work}
\label{subsec:prior_results}

We compare NNoculation with state-of-art defenses in \autoref{tab:comparision_results}, except for ABS and MNTD which are compared separately since these defenses have different goals than the others. Most prior defenses do not provide fine-grained knobs to trade-off clean accuracy (CA) against ASR, hence we did our best to optimize their performance.

In \autoref{tab:comparision_results}, we indicate, in bold font, defenses that are on the Pareto front for each BadNet (excluding ones with oracle access). We note that: (1) \sys (post-deployment) is Pareto optimal for \emph{all} BadNets, while the remaining defenses are on the Pareto front for at most two BadNets;  (2) \sys (post-deployment) is the \textbf{only defense that works consistently across all attacks.} All other defenses fail entirely for most attacks as a consequence of their restrictive assumptions. 

Quantitatively, we note that \sys reduces ASR to below $4\%$, and in most cases, to $0\%$. In return, \sys also incurs a drop in CA ranging from $1.5\%$ to $5.8\%$ with an average drop of $3.5\%$. However, we note the other defenses have even lower CA (and higher ASRs). NeuralCleanse (disregarding cases where it has oracular knowledge) has higher accuracy on GTSRB-PN and GTSRB-FSA, but also much higher ASR ($12\%$ and $28\%$ respectively).

\textbf{ABS}~\cite{absliu} relies on the assumption that a \emph{single} neuron controls the backdoor behavior. This assumption is easily circumvented by our BadNet-TCA attack that uses two triggers and activates the backdoor only when \emph{both} triggers are present. Hence, multiple neurons are required to encode backdoors, and ABS's assumptions are violated. Indeed, ABS fails to detect BadNet TCA. In contrast, \sys reduces the ASR to only $2.3\%$ on this BadNet with roughly $3.5\%$ drop in clean classification accuracy. Since the ABS executable only works on the CIFAR-10 dataset and for a specific network architecture, we could not evaluate ABS on other attack settings. 

\textbf{Meta Neural Trojan Detection (MNTD)}~\cite{mntd} is a very recent defense that like ABS flags a network as either backdoored or benign. MNTD assumes that the triggers are overlayed in pixel-space, although it does not constraint the size and shape of the trigger like NeuralCleanse and Qiao et al.~\autoref{tab:mntd} shows the detection probability of MNTD on each of the baseline BadNets. Note that while it fails on feature space attacks (FSA), as expected, it also fails on 5 other pixel-space BadNets, barely performing better than random chance. We suspect this is because MNTD is only originally evaluated on small datasets and small networks.

\begin{table}[]
\centering
\caption{Performance of MNTD on baseline BadNets \label{tab:mntd}}
\vspace{-1em}
\resizebox{\columnwidth}{!}{%
\begin{tabular}{@{}ccccccccc@{}}
\toprule
 & AAA & TCA & PN & FSA & SG & LS & MTSTA & MTMTA \\
 \midrule
\multicolumn{1}{c}{Detection Probability} & 99.9 & 0 & 100 & 59.86 & 49.37 & 56.16 & 49.64 & 50.98 \\
\bottomrule
\end{tabular}%
}
\vspace{-1em}
\end{table}

\begin{table}[ht]
\centering
\caption{Performance of NNoculation (with 3\% Threshold) on adaptive attackers during pre-deployment stage. 
\label{tab:pre_adaptive_results}}
\vspace{-1em}
\resizebox{\columnwidth}{!}{%
\begin{tabular}{cccccccc}
\toprule
\multicolumn{1}{l}{} & \multicolumn{2}{c}{BadNet (Baseline)} & \multicolumn{2}{c}{\sys (pre-)} & \multicolumn{2}{c}{\sys (post-)}  \\
BadNet & CA & ASR & CA & ASR & CA & ASR \\
\cmidrule(lr){1-1} \cmidrule(lr){2-3} \cmidrule(lr){4-5} \cmidrule(lr){6-7} 
MNIST-AAA & 96.74 & 88.82 & 87.38 & 1.3 & 93.22 & 0 \\
\midrule
CIFAR10-TCA & 83.65 & 99.94 & 81.51 & 25.51 & 80.14 & 1.88 \\
\midrule
GTSRB-PN & 95.36 & 99.93 & 94.62 & 0 & 92.97 & 0 \\
GTSRB-FSA & 93.47 & 94.15 & 91.06 & 3.11 & 90.07 & 3.48 \\
\midrule
YouTube-SG & 97.99 & 99.95 & 94.33 & 19.8 & 93.43 & 0 \\
YouTube-LS & 96.13 & 91.68 & 93.82 & 14.64 & 93.51 & 0 \\
YouTube-MTSTA & 90.87 & \{94.2, 94.4, 100\} & 86.81 & \{0, 0, 0\} & 86.25 & \{0, 0, 0\}\\
YouTube-MTMTA & 91.09 & \{91.7, 92.2, 100\} & 87.96 & \{0, 0, 0\} & 87.79 & \{0, 0, 0\} \\
\bottomrule
\end{tabular}
}
\vspace{-1em}
\end{table}

\begin{table}[ht]
\centering
\caption{\sys under adaptive online attack.  \label{tab:multi_training}}
\vspace{-1em}
\resizebox{\columnwidth}{!}{%
\begin{tabular}{@{}ccccc@{}}
\toprule
 &  & NNoculation (pre-) & \multicolumn{2}{c}{NNoculation (post-)} \\
 &  & (ls, eb, sg) & Stage 1 (ls, eb) & Stage 2 (sg) \\ 
\midrule
\multirow{2}{*}{BadNet-MTMTA} & CA & 92.13 & 91.51 & 91.31 \\
 & ASR & 30, 7.12, 13.77 & 0, 0, 6.67 & 0, 0, 0 \\ \bottomrule
\end{tabular}%
}
\vspace{-1em}
\end{table}

\subsection{Performance Under Adaptive Attacks}
\label{adapt_attacks}

Standard security practice suggests that defenses should withstand adaptive attacks that are aware of the defense. We show that \sys is immune to two such attacks and leave the exploration of more advanced adaptive attacks for future work. 

\textbf{Adaptive attacks against pre-deployment defense.} Adaptive attackers could craft more robust BadNets by  adding noise to training data themselves, emulating one step of our defense. However, intuitively, this attack will fail because retraining with higher learning rates always provides the defender with a possibility to reduce ASR (see~\autoref{fig:pre-deployment-ca-asr}). Our evaluation supports our intuition --- ~\autoref{tab:pre_adaptive_results} shows that \sys under adaptive attack is competitive with baseline \sys (both post-deployment) on all the BadNets.

\textbf{Dealing With Adaptive Online Attackers.} An adaptive attacker could train a BadNet with multiple triggers (such as BadNet-MTMTA) and deploy triggers sequentially, for example, use lipstick and eyebrow triggers first and then use the sunglasses trigger once \sys retrains for the first two. However, because \sys adapts in the field to new attacks, it reduces the ASR for lipstick and eyebrow triggers to zero after the first round of re-training, and that of the sunglasses trigger also to zero after the subsequent round (see~\autoref{tab:multi_training}).

\subsection{Sensitivity to Dataset Size}

We address the performance of \sys on large real-world datasets and impact of reduced access to validation data for the defender. 

\begin{table*}[]
\centering
\caption{Sensitivity of NNoculation (with 3\% Threshold) towards different validation dataset sizes on simple baseline BadNets.\label{tab:less_data}}
\vspace{-1em}
\resizebox{\textwidth}{!}{%
\begin{tabular}{ccccccccccccccc}
\toprule
\multicolumn{1}{l}{} & \multicolumn{1}{l}{} & \multicolumn{1}{l}{} & \multicolumn{4}{c}{25\% of validation data} & \multicolumn{4}{c}{50\% of validation data} & \multicolumn{4}{c}{75\% of validation data} \\
\cmidrule{1-1} \cmidrule(lr){2-3} \cmidrule(lr){4-7} \cmidrule(lr){8-11} \cmidrule(lr){12-15}
\multicolumn{1}{l}{} & \multicolumn{2}{c}{BadNet (Baseline)} & \multicolumn{2}{c}{\sys (pre-)} & \multicolumn{2}{c}{\sys (post-)} & \multicolumn{2}{c}{\sys (pre-)} & \multicolumn{2}{c}{\sys (post-)} & \multicolumn{2}{c}{\sys (pre-)} & \multicolumn{2}{c}{\sys (post-)} \\
BadNet & CA & ASR & CA & ASR & CA & ASR & CA & ASR & CA & ASR & CA & ASR & CA & ASR \\
\cmidrule{1-1} \cmidrule(lr){2-3} \cmidrule(lr){4-5} \cmidrule(lr){6-7} \cmidrule(lr){8-9} \cmidrule(lr){10-11} \cmidrule(lr){12-13} \cmidrule(lr){14-15}
SG & 97.89 & 99.98 & 95.23 & 87.37 & 85.58 & 33.71 & 95.08 & 72.29 & 87.6 & 11.12 & 94.39 & 61.73 & 90.05 & 2.31 \\
LS & 97.19 & 91.51 & 94.07 & 70.46 & 88.41 & 3.84 & 93.86 & 58.63 & 89.3 & 2.5 & 94.66 & 45.87 & 92.45 & 0 \\
PN & 97.19 & 91.51 & 91.89 & 36.79 & 91.09 & 5.19 & 93.42 & 0 & 93.32 & 0 & 94.1 & 0 & 91.27 & 0 \\
IN & 71.88 & 100 & 68.1 & 65.86 & 66.47 & 0 & 67.45 & 38.63 & 67 & 0 & 68.07 & 22.86 & 67.84 & 0 \\
\bottomrule
\end{tabular}
}
\vspace{-1em}
\end{table*}

\textbf{\sys on Imagenet.} We evaluated \sys on BadNet-IN (DenseNet-121 with ImageNet) to verify the scalability of our defense to large datasets and complex network architectures. Our pre-deployment defense has a classification accuracy (CA) of $69.14\%$ (down from $71.88\%$) and attack success rate (ASR) of $43.45\%$. Using our post-deployment defense reduces ASR to $0\%$, while maintaining the CA at $68.27\%$. 

\textbf{Impact of validation dataset size.} Like other defenses, \sys relies to some extent on the availability of clean validation data. In our experiments, we have used standard validation dataset sizes to evaluate \sys and prior work. The validation set for YouTube Face is already tiny --- it has only 9 images per class, and training a network from scratch on validation data alone would result in large $17\%$ drop in accuracy on this dataset.

Nonetheless, to understand the performance of \sys with even smaller validation datasets, we evaluate \sys on BadNet-SG, BadNet-LS, BadNet-PN and BadNet-IN with 25\%, 50\% and 75\% of the original validation dataset and report the results in~\autoref{tab:less_data}. \sys still has relatively low ASR in all cases except on BadNet-SG with $25\%$ of the original validation data, where the ASR increases to $33\%$. Note, however, that $25\%$ of the original validation data represents only 2 samples per class for the YouTube Face dataset. Notably, on BadNet-IN, \sys's ASR is still $0\%$ even with $25\%$ of the original validation data.

\section{Discussion}

\subsection{Performance on Additional Attacks}
\paragraph{Clean Label Attack (CLA)~\cite{cla_kang}} During training, the attacker only poisons the target class with the trigger. During test time, any input with the trigger will be classified as the target label. BadNet-CLA is trained on MNIST dataset using a fixed noise pattern as the trigger and $\mathcal{T} = 0$. To explore the clean label attack, we devised a custom network architecture as described in~\autoref{tab:cla_architecture}. \sys succeeds in defending against BadNet-CLA as shown in~\autoref{tab:cla}.

\begin{table}[]
\centering
\caption{DNN Architecture for Clean-Attack (CLA) BadNet}
\vspace{-1em}
\resizebox{\columnwidth}{!}{%
\begin{tabular}{ccccc}
\toprule
Layer Type & \# of Channels & Filter Size & Stride & Activation \\
\midrule
Conv & 16 & $5\times5$ & 1 & ReLU \\
MaxPool & 16 & $2\times2$ & 2 & - \\
Conv & 4 & $5\times5$ & 1 & ReLU \\
MaxPool & 4 & $2\times2$ & 2 & - \\
FC & 512 & - & - & ReLU \\
FC & 10 & - & - & Softmax \\
\bottomrule
\end{tabular}
}
\label{tab:cla_architecture}
\vspace{-1em}
\end{table}

\begin{table}[]
\small
\centering
    \caption{Performance on BadNet-CLA}
    \vspace{-1em}
\begin{tabular}{@{}lcccccc@{}}
\toprule
 \multicolumn{2}{c}{Baseline BadNet} & \multicolumn{2}{c}{\sys (pre-)} & \multicolumn{2}{c}{\sys (post-)} \\ 
 CA & ASR & CA & ASR & CA & ASR \\
 \cmidrule(lr){1-2} \cmidrule(lr){3-4} \cmidrule(lr){5-6}
 89.02 & 100 & 91.18 & 2.67 & 95.34 & 0 \\
 \bottomrule
\end{tabular}%
\label{tab:cla}
\vspace{-1em}
\end{table}

\paragraph{Imperceptible Attack~\cite{nguyen2021wanet}} The attacker uses an invisible trigger proposed by~\cite{nguyen2021wanet} to generate the imperceptible backdoor attack. To evaluate against this attack, we trained two BadNets on MNIST and GTSRB datasets using network architectures proposed in~\cite{strip2019acsac} and~\cite{neuralcleanse}, respectively. The target label for both these attacks is $\mathcal{T} = 0$. We evaluated these attacks and show in~\autoref{tab:imperceptible_trigger}, that NNoculation defends against both. As noted in~\cite{nguyen2021wanet}, we verified that Pruning, STRIP, and NeuralCleanse fail against this imperceptible attack.

\begin{table}[h]
\caption{Performance on imperceptible attacks.}
\vspace{-1em}
\label{tab:imperceptible_trigger}
\resizebox{\columnwidth}{!}{%
\begin{tabular}{cccccccc}
\toprule
 & \multicolumn{2}{c}{BadNet (Baseline)} & \multicolumn{2}{c}{NNoculation (pre-)} & \multicolumn{2}{c}{NNoculation (post-)} \\
Dataset & CA    & ASR   & CA    & ASR & CA    & ASR \\
\cmidrule(lr){1-1} \cmidrule(lr){2-3} \cmidrule(lr){4-5} \cmidrule(lr){6-7} 
MNIST  & 98.66 & 99.87 & 93.5 & 0 & 95.13 & 0 \\
  \midrule
GTSRB & 94.29 & 94.53 & 91.56 & 0   & 90.92 & 0  \\
\bottomrule
\end{tabular}%
}
\vspace{-1em}
\end{table}

\subsection{Ablations} 
\paragraph{Pre-deployment Defense}
To evaluate the effectiveness of noise augmentation ($A$) and adaptive learning rate (Adapt $\alpha$), we evaluate \sys on BadNet-SG, BadNet-LS and BadNet-PN by conducting an ablation study on the pre-deployment defense, and report the results in~\autoref{tab:ablation}. \sys achieves 0\% ASR in all cases, however, on BadNet-LS, \sys w/o \{Adapt $\alpha$\} achieves 1.86\% higher CA compared to \sys. However, \sys w/o \{Adapt $\alpha$\} would fail on adaptive attacks against pre-deployment defense as noted in~\autoref{adapt_attacks}. 

\begin{table}[]
\caption{Ablation study on pre-deployment defense.}
\vspace{-1em}
\label{tab:ablation}
\resizebox{\columnwidth}{!}{%
\begin{tabular}{cccccccc}
\toprule
 &  & \multicolumn{2}{c}{BadNet (Baseline)} & \multicolumn{2}{c}{NNoculation (pre-)} & \multicolumn{2}{c}{NNoculation (post-)} \\
          BadNet          &          Method            & CA                     & ASR                    & CA    & ASR   & CA    & ASR   \\

\cmidrule(lr){1-1} \cmidrule(lr){2-2} \cmidrule(lr){3-4} \cmidrule(lr){5-6} \cmidrule(lr){7-8} 
                    
\multirow{4}{*}{SG} & NNoc. (Proposed)                   & \multirow{4}{*}{97.89} & \multirow{4}{*}{99.98} & \textbf{94.18} & \textbf{35.22} & \textbf{92.03} & \textbf{0}     \\
                    & NNoc. w/o \{$A$\}                &                        &                        & 95.63 & 75.29 & 92.19 & 1.01  \\
                    & NNoc. w/o \{$A$, Adapt $\alpha$\}      &                        &                        & 97.62 & 98.14 & 85.09 & 4.82  \\
                    & NNoc. w/o \{Adapt $\alpha$\}          &                        &                        & 94.93 & 71.71 & 92.67 & 1.26  \\
\midrule
\multirow{4}{*}{LS} & NNoc. (Proposed)                    & \multirow{4}{*}{97.19} & \multirow{4}{*}{91.51} & 93.99 & 32.34 & 93.15 & 0  \\
                    & NNoc. w/o \{$A$\}                &                        &                        & 94.86 & 74.49 & 92.83 & 24.17 \\
                    & NNoc. w/o \{$A$, Adapt $\alpha$\}      &                        &                        & 97.97 & 89.91 & 89.91 & 14.52 \\
                    & NNoc. w/o \{Adapt $\alpha$\}          &                        &                        & \textbf{94.25} & \textbf{24.3}  & \textbf{95.01} & \textbf{0}  \\
\midrule
\multirow{4}{*}{PN} & NNoc. (Proposed)                    & \multirow{4}{*}{95.46} & \multirow{4}{*}{99.82} & \textbf{92.9}  & \textbf{13.37} & \textbf{93.01} & \textbf{0}     \\
                    & NNoc. w/o \{$A$\}                &                        &                        & 94.3  & 36.79 & 91.66 & 0     \\
                    & NNoc. w/o \{$A$, Adapt $\alpha$\}      &                        &                        & 96.02 & 98.34 & 94.21 & 74.54 \\
                    & NNoc. w/o \{Adapt $\alpha$\}          &                        &                        & 94.74 & 38.73 & 92.68 & 0     \\
\bottomrule
\end{tabular}%
}
\vspace{-1em}
\end{table}

\paragraph{Post-deployment Defense} For identifying and reverse-engineering triggers, we tried simpler alternatives, including averaging quarantined images and subtracting from the average of clean validation images. We found that averaging/subtraction works well if triggers are additive and at fixed locations since the backdoor stands out. However, we found this worked poorly for variable-location triggers. Since our primary goal is to avoid making any assumptions on the \texttt{poison}($x$) function, we thus moved to the CycleGAN approach which, in theory (i.e., with a sufficiently complex network and sufficient data) should be able to learn any poisoning function. Note that the CycleGAN is a strict generalization of the simple averaging approach aforementioned.

\subsection{CycleGAN Output Quality vs. Training Data} \label{app-sec:cyclegan_quality_training_data}
\begin{figure}[ht]
    \centering
    \includegraphics[width=\columnwidth]{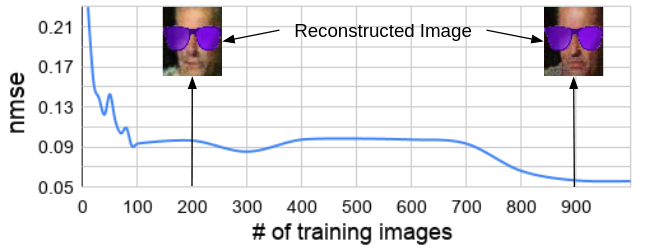}
    \vspace{-2em}
    \caption{The normalized mean square error (nmse) between reconstructed and actual triggers, and examples of reconstructed backdoors (in inset), vs. amount of training data.}
    
    \vspace{-1em}
    \label{fig:quality}
\end{figure}

Our results are consistent with the original CycleGAN paper~\cite{CycleGAN2017} that achieved high-quality results on small datasets with as few as 400 images, for example, on the facades $\leftrightarrow$ photographs and artistic style transfer tasks, especially where domains are similar. Likewise, we show in~\autoref{fig:quality} that we can recover relatively high quality trigger reconstructions (see inset, and~\autoref{fig:cyclegan_rev_triggers}) with only 200 images in each domain (we use 500 clean validation images in domain 1 and 200 quarantined images in domain 2 in our implementation). Although the reconstruction mse is even lower with 1000 images, empirically, as also observed in prior work~\cite{neuralcleanse}, exact trigger reconstruction is not needed for backdoor unlearning.
 
\subsection{CycleGAN Output Quality vs. Poison/Clean Data Ratio in Online Stream of Test Inputs}
\begin{figure}[]
    \begin{center}$
    \begin{array}{ccccccc}
    
    \includegraphics[width=0.4in]{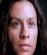}&
    \includegraphics[width=0.4in]{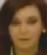}&
    \includegraphics[width=0.4in]{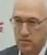}&
    \includegraphics[width=0.4in]{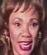}&
    \includegraphics[width=0.4in]{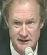}&
    \includegraphics[width=0.4in]{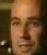}\\
    
    \includegraphics[width=0.4in]{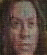}&
    \includegraphics[width=0.4in]{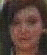}&
    \includegraphics[width=0.4in]{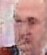}&
    \includegraphics[width=0.4in]{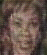}&
    \includegraphics[width=0.4in]{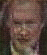}&
    \includegraphics[width=0.4in]{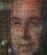}\\
    
    \includegraphics[width=0.4in]{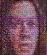}&
    \includegraphics[width=0.4in]{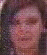}&
    \includegraphics[width=0.4in]{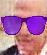}&
    \includegraphics[width=0.4in]{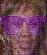}&
    \includegraphics[width=0.4in]{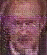}&
    \includegraphics[width=0.4in]{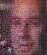}\\
    
    \includegraphics[width=0.4in]{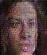}&
    \includegraphics[width=0.4in]{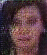}&
    \includegraphics[width=0.4in]{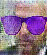}&
    \includegraphics[width=0.4in]{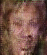}&
    \includegraphics[width=0.4in]{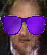}&
    \includegraphics[width=0.4in]{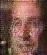}\\
    
    \includegraphics[width=0.4in]{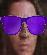}&
    \includegraphics[width=0.4in]{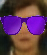}&
    \includegraphics[width=0.4in]{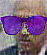}&
    \includegraphics[width=0.4in]{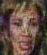}&
    \includegraphics[width=0.4in]{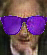}&
    \includegraphics[width=0.4in]{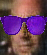}\\
    
    \end{array}$
    \end{center}
    \vspace{-1.5em}
    \caption{Examples of synthetic poisoned samples for post-deployment treatment generated by CycleGAN approximation of \texttt{poison}($x$) for BadNet-SG. Top row: clean images;  Remaining rows: synthetic data produced by CycleGAN trained on quarantined data collected using poison/clean data ratio of 0.02, 0.06, 0.1, 0.5, respectively.}
    \label{fig:cycleGAN-ratio-visual_appendix}
    \vspace{-2em}
    \end{figure}

To qualitatively understand \sys's post-deployment defense,~\autoref{fig:cycleGAN-ratio-visual_appendix} shows a selection of backdoor images generated by the CycleGAN for BadNet-SG. Recall these are generated by feeding clean validation data into the CycleGAN's generator. Note that we begin to see good trigger insertion after training the CycleGAN on quarantined data collected from a 6\% of poisoned images in $\mathcal{D}_{stream}$. As the CycleGAN is trained on more poison data in the quarantined data, the trigger insertion becomes more reliable (the last row of~\autoref{fig:cycleGAN-ratio-visual_appendix}).

\subsection{Implications for Future Defenses} Empirical observations from our pre- and post-deployment defenses have important implications for future defenses. 
First, the fact that re-training with \emph{random} perturbations is about as effective as unlearning with a targeted search for backdoors demonstrates the potential futility of the latter (or conversely, the need to significantly improve backdoor search mechanisms). Second, we note that our post-deployment defense is \emph{complementary} to any pre-deployment defense, especially since we show the efficacy of our post-deployment defense even if the pre-deployment does not significantly reduce ASR.  

\subsection{Limitations and Threats to Validity} \sys has been evaluated only in the context of BadNet attacks in the image domain. Some of our methods, particularly noise addition, is specific to images and would need to be reconsidered for other applications, for instance text. Further, our attack model is restricted to training data poisoning as an attack strategy; one could imagine attackers that make custom changes to the weights of a trained BadNet to further evade defenses. Finally, we have assumed a computationally capable defender (although one that lacks access to high-quality training data); one can imagine a setting where the defender has only limited computational capabilities and cannot, for instance, train a CycleGAN. Defenses such as fine-pruning are more appropriate in that setting, but do not currently appear to work across a broad spectrum of attacks. Neural Cleanse and ABS, on the other hand, have relatively high computational costs. 

\subsection{Run-time Overhead Vs. Complexity}
STRIP runs 1000 images through the network at run-time per test image, resulting in a 1000x reduction in run-time throughput. Both NeuralCleanse and ABS take upto an hour during pre-deployment and nonetheless fail in several cases. In comparison, NNoculation's pre-deployment defense is quick (tens of minutes), and our online CycleGAN runs in the background (i.e., in parallel) with the deployed network, and thus does not add throughput cost. NNoculation's primary computational cost is in CycleGAN training (takes roughly 5 hours), but this can happen while the deployed network continues to quarantine suspicious inputs. Note that this computational effort is amortized over the network’s lifetime, yielding the only defense that works against a range of attacks; and can be further reduced in the future using advanced training methods, but these were not the focus of this work. 

\section{Related Works}
\label{sec:rel}

\textbf{Attacks}
There are two broad classifications of attacks on machine learning~\cite{biggio_wild_2018}, inference-time attacks (i.e., those that make use of adversarial perturbations~\cite{liu2019adversarial, szegedy_intriguing_2013, GoodfellowSS14}) and training-time attacks, as we explore in this work. BadNets~\cite{badnets} proposed the first backdoor attack on DNNs through malicious training with poisoned data, showcasing both targeted and random attacks where an attacker aims to force a backdoored DNN to misclassify inputs with a specific trigger as the target label (targeted attacks) or a random label (random attacks) in the context of pretrained online models and transfer learning setting. There are two ways in which a DNN can be backdoored: dirty-label attacks where training data is mislabelled (such as those in~\cite{neuraltrojans,sunglassesattack}), and clean-label attacks, where training data is cleanly labeled, as in Poison Frogs~\cite{poisonfrogs}. 
    
\textbf{Defenses}
We have already discussed and compared \sys with several state-of-art defenses against backdoored neural networks~\cite{badnets, neuraltrojans, cla_kang, sunglassesattack, poisonfrogs, nguyen2021wanet}. We note that defenses against training time attacks on DNNs under different threat models have also been considered, including defenses that assume the defender has access to \emph{both} clean and backdoored inputs pre-deployment~\cite{tran2018spectral,steinhardt2017certified, certified_defense} (i.e., data poisoning attacks that, as discussed in~\autoref{sec:bckgnd}, are weaker than the backdooring threat that we consider), defenses against backdoored federated learning~\cite{blanchard2017machine,sun2019really}, and defenses for simpler regression models~\cite{8418594}. However, none of these methods directly translate to our threat model.
\section{Conclusion}
\label{sec:cncl}

We proposed a novel two-stage Neural Network inoculation ({\bf NNoculation}) defense against backdoored neural networks (BadNets). In the pre-deployment stage, we use noise-augmentation and high learning rates to activate a broad-spectrum of BadNet neurons, allowing the neurons to be fine-tuned with clean validation data, and alleviating the need for unrealistic assumptions on trigger characteristics. Post-deployment, we quarantine data suspected of containing of backdoors and use a CycleGAN to reverse engineer the attacker's poisoning function. This knowledge is then used for targeted retraining of the BadNet to further reduce the attack success rate.  Our experiments comparing NNoculation to prior  defenses show that it is the only defense that works across a range of backdoor attacks, while all other methods fail on one or multiple attacks. 

\bibliographystyle{ACM-Reference-Format}
\bibliography{sample-base}

\appendix

\section{Appendix \label{sec:appendix}}

\subsection{Further Study of Pre-deployment} \label{app-sec:hyper}

To evaluate the sensitivity of pre-deployment against training hyperparameters, we prepare additional BadNet variants by modifying different training hyperparameters during BadNet training. These settings are presented in Tables~\ref{tab:variant-settings-SG}, and~\ref{tab:variant-settings-PN}. We use the heuristic described in~\autoref{sec:exps} to select $\theta_{aug}$ for each BadNet, and report the change in CA and change in ASR in~\autoref{tab:hyperparameter-variant}. We find that in most cases, the CA degradation is $\approx 3\%$ as desired, with ASR reduction ranging from $-66.5\%$ to $-100\%$; i.e., in some settings, the pre-deployment defense is able to remove the backdoor behavior entirely. These results appear to support the idea that our pre-deployment defense method is broadly applicable in spite of varying attacker BadNet training hyperparameters.

\begin{figure*}[ht]
\centering
\subfloat[BadNet-SG CA]{\label{fig:sg-acc}\includegraphics[width=0.325\textwidth]{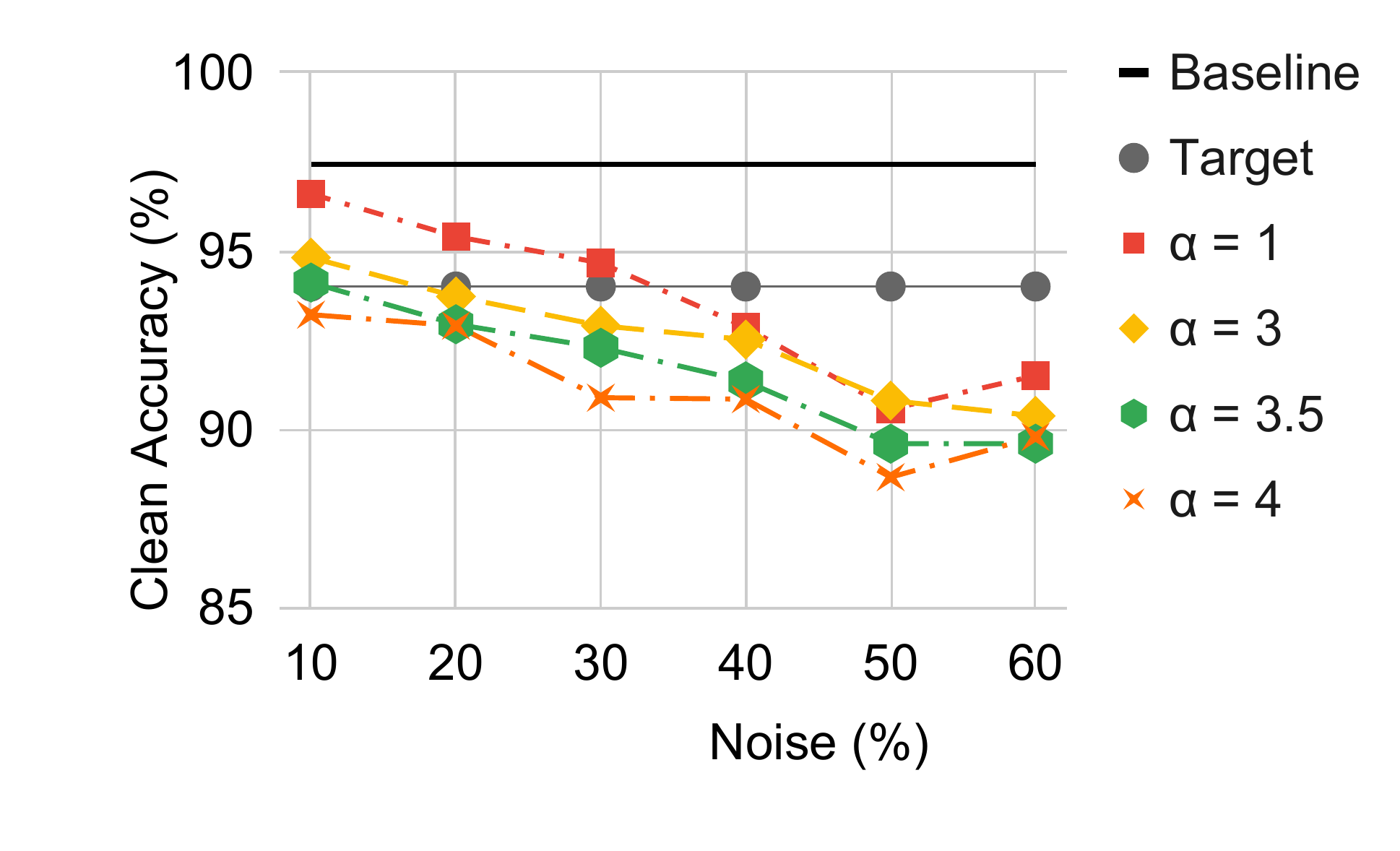}} \hfill
\subfloat[BadNet-LS CA]{\label{fig:ls-acc}\includegraphics[width=0.325\textwidth]{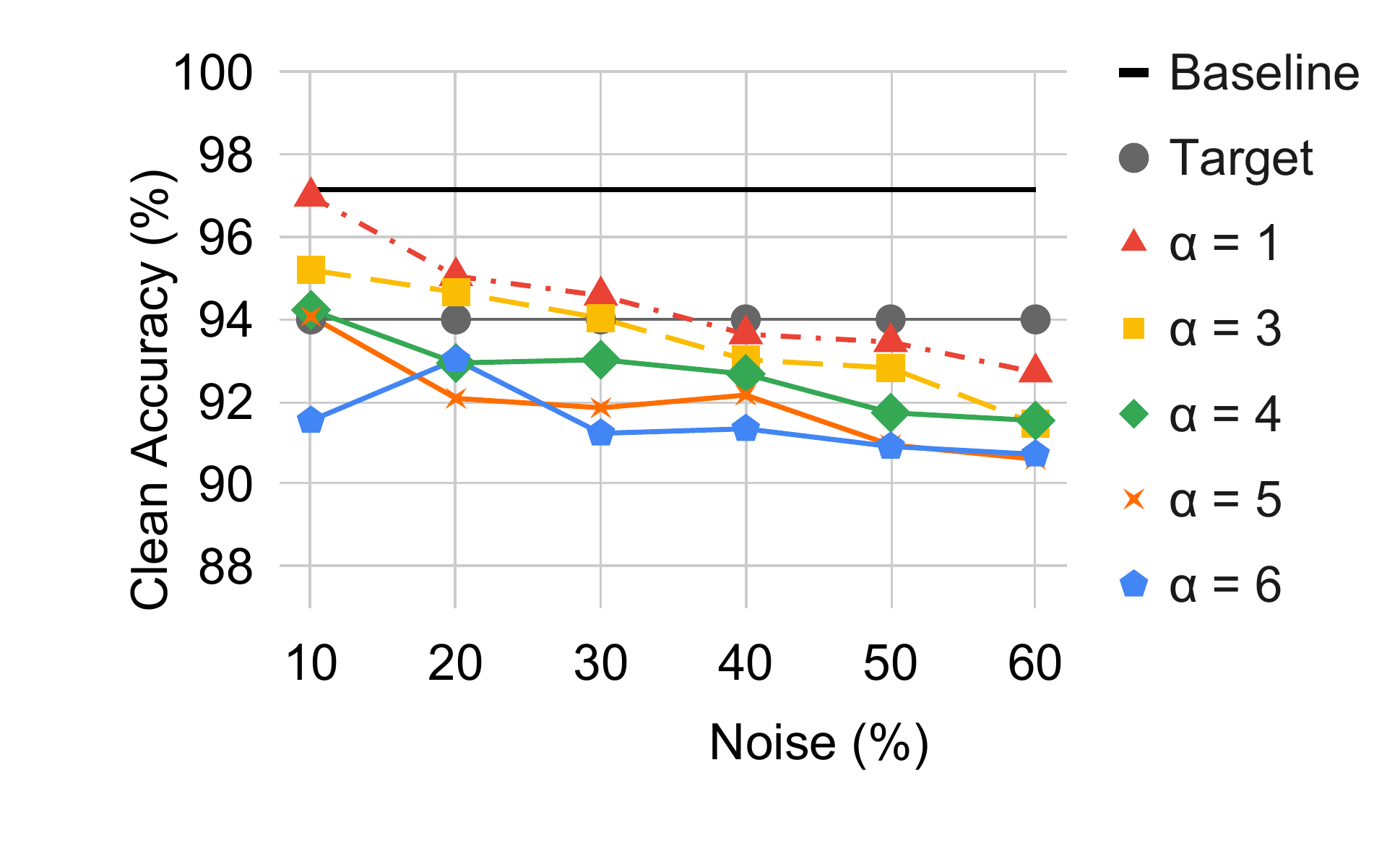}}  \hfill
\subfloat[BadNet-PN CA]{\label{fig:gt-acc}\includegraphics[width=0.325\textwidth]{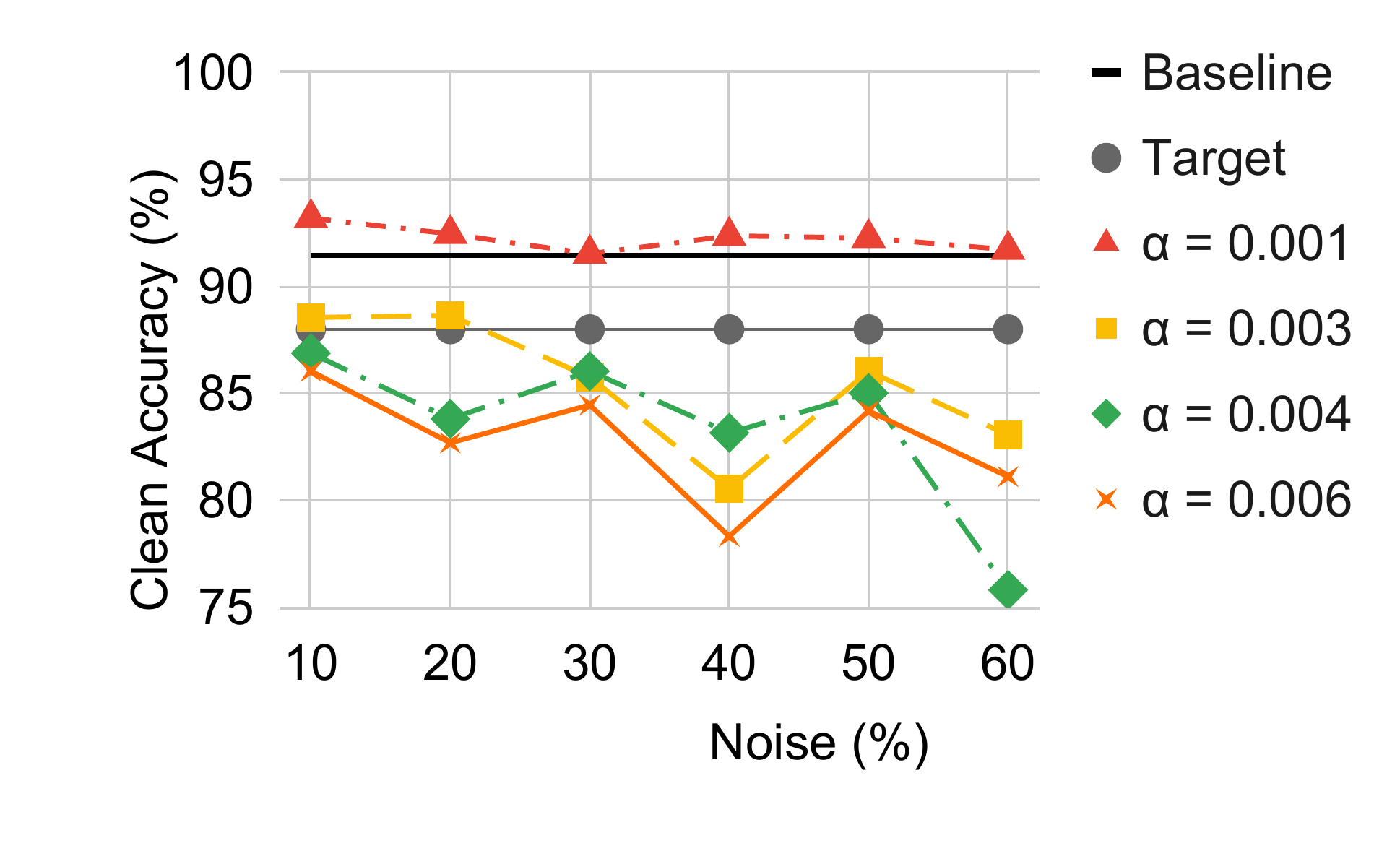}} \\
\subfloat[BadNet-SG ASR]{\label{fig:sg-as}\includegraphics[width=0.325\textwidth]{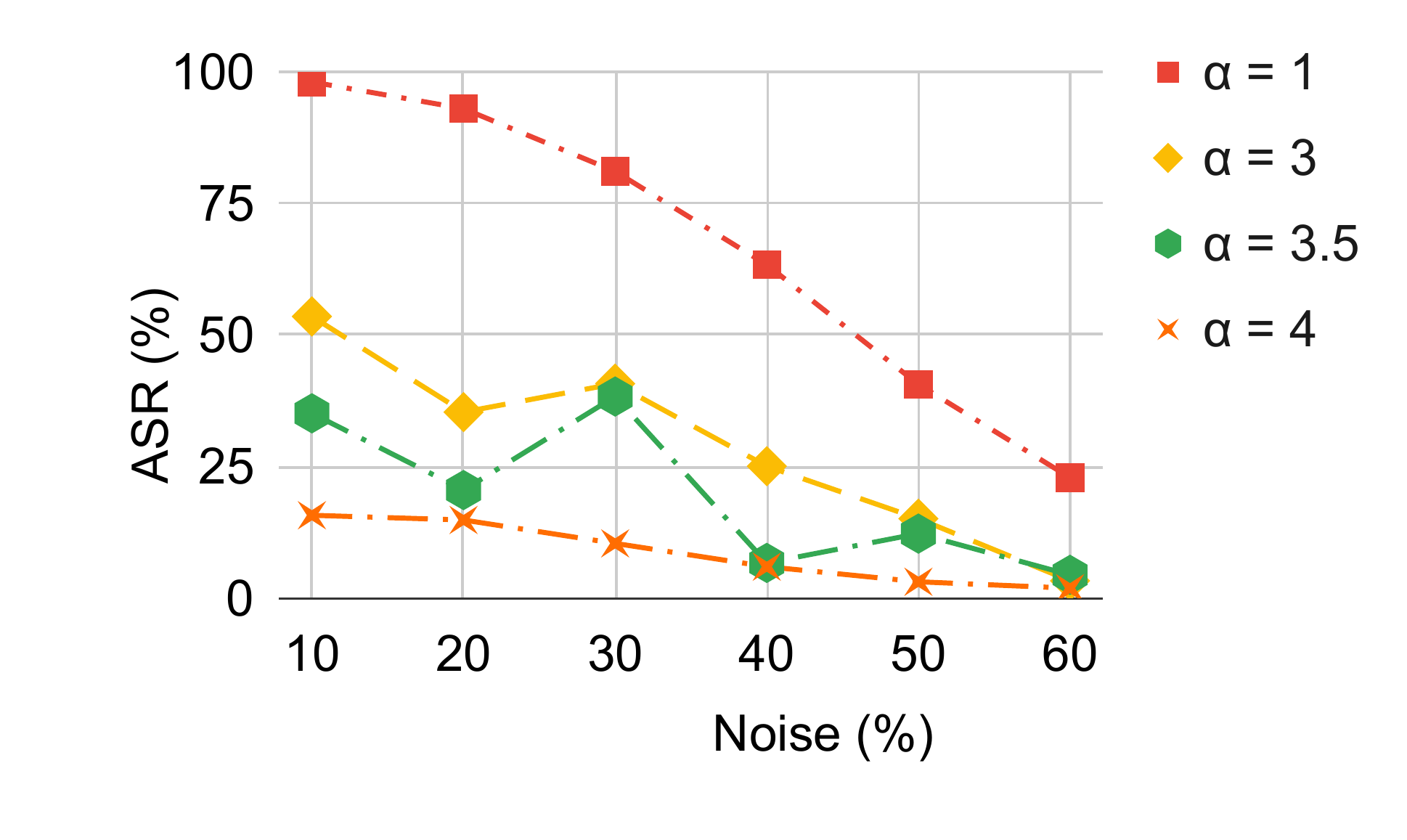}} \hfill
\subfloat[BadNet-LS ASR]{\label{fig:ls-asr}\includegraphics[width=0.325\textwidth]{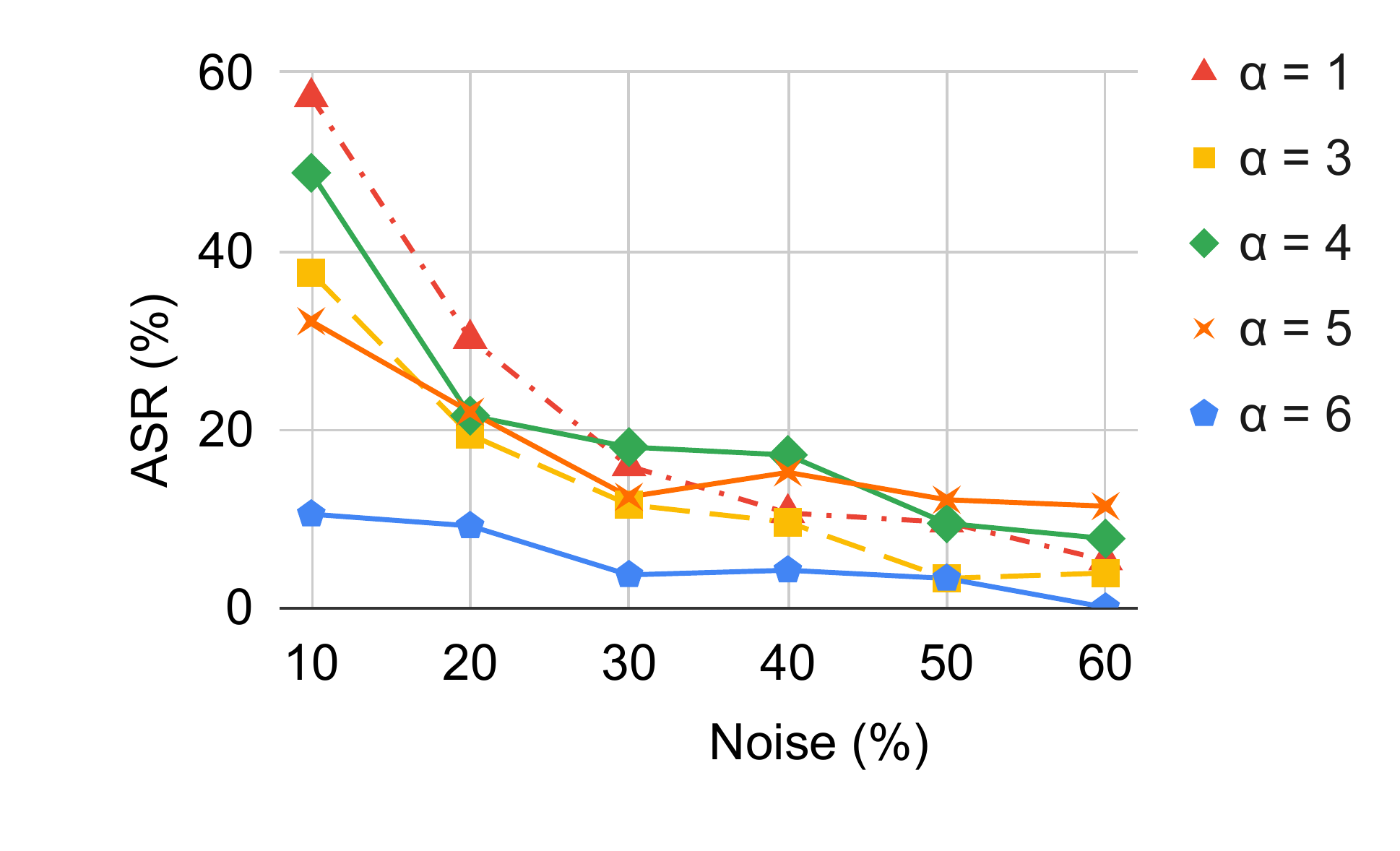}}  \hfill
\subfloat[BadNet-PN ASR]{\label{fig:gt-asr}\includegraphics[width=0.325\textwidth]{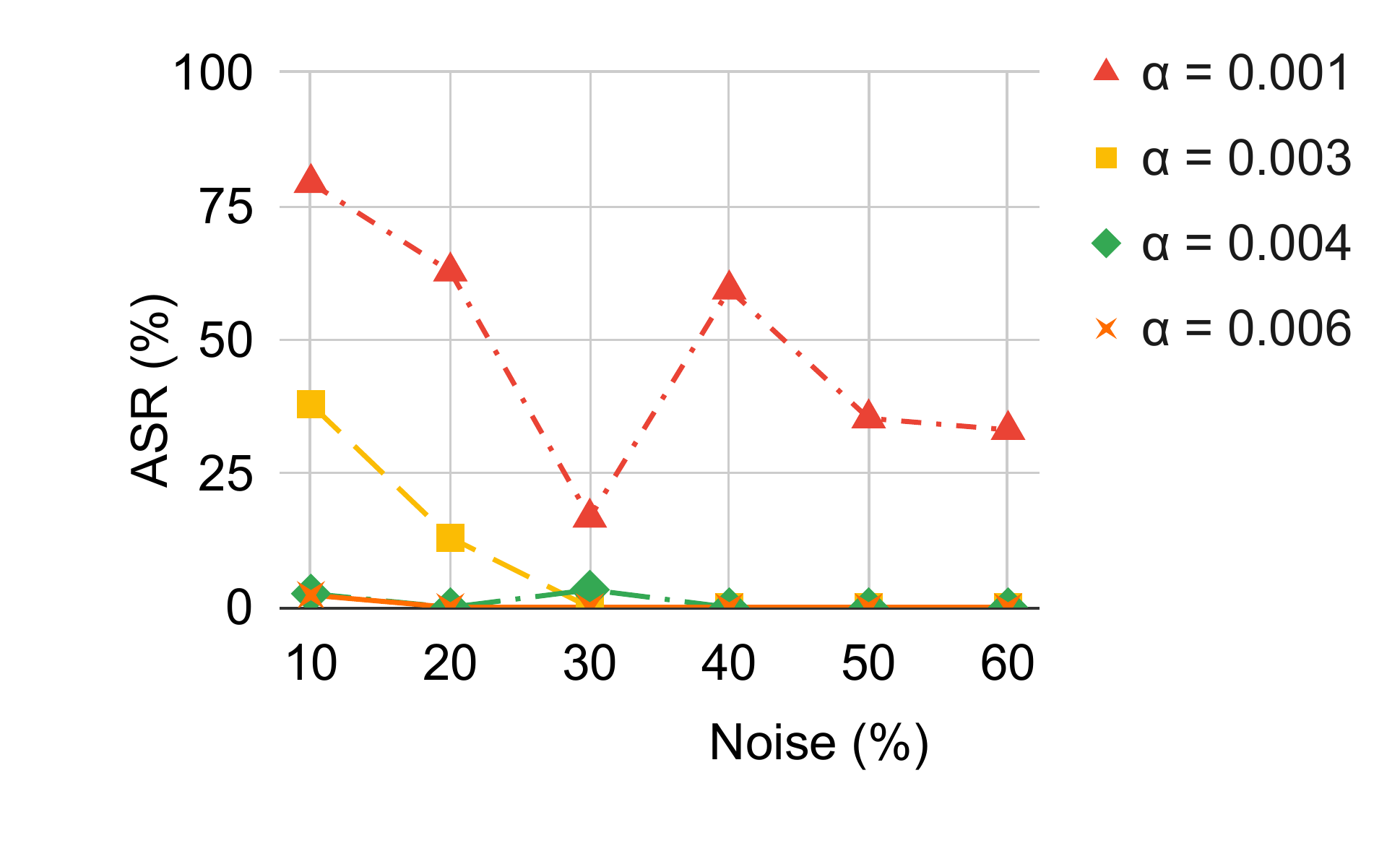}}
\caption{Effect of pre-deployment treatment on CA (on evaluation data) and ASR (on test data) under varying $\alpha$ and $\gamma$ settings.%
\label{fig:appendix_pre-deployment-ca-asr}}
\vspace{-1.5em}
\end{figure*}
   
\begin{table}[h]
\centering
\caption{Hyperparameter variants for BadNet-SG.}
\vspace{-1em}
 \resizebox{\columnwidth}{!}{%
\begin{tabular}{@{}lcccc@{}}
\toprule
 & \multicolumn{4}{c}{Variants} \\
hyperparameter & \texttt{ORIG} & \texttt{BATCH} & \texttt{ATK Type} & \texttt{~\cite{finepruning}} \\ \midrule
batch size & 1283 & 256 & 1283 & 1283 \\
epochs & 200 & 200 & 200 & 200 \\
learning rate & 1 & 1 & 1 & 0.001 \\
optimizer & ADADELTA & ADADELTA & ADADELTA & Adam \\
preprocessing & divide by 255 & divide by 255 & divide by 255 & raw \\
attack type & 2-step & 2-step & 1-step & 2-step \\ \bottomrule
\end{tabular}
}
\label{tab:variant-settings-SG}
\vspace{-1em}
\end{table}

\begin{table}[h]
\centering
\caption{Hyper-parameter variants for BadNet-PN.}
\vspace{-1em}
 \resizebox{\columnwidth}{!}{%
\begin{tabular}{@{}lccccc@{}}
\toprule
 & \multicolumn{4}{c}{Variants} \\
hyperparameter & \texttt{ORIG} & \texttt{RAW} & \texttt{LR} & \texttt{SGD} \\ \midrule
batch size & 32 & 32 & 32 & 32 \\
epochs & 15 & 15 & 15 & 15 \\
learning rate & 0.001 & 0.001 & 0.003 & 0.01 \\
optimizer & adam & adam & adam & SGD \\
preprocessing & divide by 255 & raw & divide by 255 & divide by 255 \\
attack type & 2-step & 2-step & 2-step & 2-step \\ \bottomrule
\end{tabular}
}
\label{tab:variant-settings-PN}
\vspace{-1em}
\end{table}
    
\begin{table}[h]
    \centering
    \caption{Pre-deployment defense of BadNet-SG, PN variants.}
    \vspace{-1em}
    \resizebox{\columnwidth}{!}{%
    \begin{tabular}{@{}lcccccc@{}}
    \toprule
     & \multicolumn{3}{c}{$\theta_{aug}$ for BadNet-SG} & \multicolumn{3}{c}{$\theta_{aug}$ for BadNet-PN} \\ 
     & \texttt{BATCH} & \texttt{ATK} & \texttt{~\cite{finepruning}} & \texttt{RAW} & \texttt{LR} & \texttt{SGD} \\
     \cmidrule(lr){2-4} \cmidrule(lr){5-7}
    CA change (\%) & $-$3.54 & $-$3.8 & $-$3.7 & $-$3.5 & $-$1.7 & $-$0.3  \\
    ASR change (\%) & $-$90.35 & $-$93.0 & $-$69.5 & $-$66.0 & $-$87.8 & $-$100 \\ \bottomrule
    \end{tabular}%
    }
    \label{tab:hyperparameter-variant}
    \vspace{-1em}
\end{table}
    
\subsection{Comparision with Neural Attention Distillation (NAD)}
(NAD) is a very recent backdoor defense proposed by Li et al.~\cite{li2021neural}. We observe that it suffers a 9\% drop in CA and yet has an high ASR of 75.66\% and 19.13\% on lipstick and sunglasses triggers respectively for YouTube-MTMTA. NAD is also ineffective on BadNet CIFAR10-TCA with an ASR = 99\%.

\subsection{CycleGAN Details} \label{app-sec:cyclegan_details_params}

To train the CycleGAN during post-deployment defense, we use the same architecture and training hyper-parameters from the original CycleGAN paper~\cite{CycleGAN2017} which demonstrated impressive unpaired image-to-image translation for various tasks. We adopt CycleGAN-Keras implementation from~\cite{cyclegan_implementation} repository. The generator architecture, in~\autoref{tab:generator}, has 3 convolutional layers, 9 residual blocks, 2 fractionally strided and one convolutional layer. The discriminator, in~\autoref{tab:discrminator}, uses 70×70 PatchGANs~\cite{patchgan} and produces a 1-dimensional output.

We train all the networks from scratch for 200 epochs using Adam optimizer~\cite{adam_optim}. The learning rate is held constant at 0.0002 for first 100 epochs and then lineary decay to zero for the last 100 epochs. During optimizing the discriminator, the objective is halved to ensure that the discriminator learns slowly relative to the generator.

\begin{table}[h]
\caption{CycleGAN discriminator architecture~\cite{CycleGAN2017}. Kernel size is 4$\times$4, padding is `Same' and LeakyReLU, $\alpha=2$.}
\vspace{-1em}
\label{tab:discrminator}
\resizebox{\columnwidth}{!}{%
\begin{tabular}{cccccc}
\toprule
Layer Name & Filters  & Stride & Normalization & Activation \\
\midrule
c64        & 64         & 2      & False         & LeakyReLU  \\
c128       & 128        & 2      &  True          & LeakyReLU  \\
c256       & 256        & 2      &  True          & LeakyReLU  \\
c512       & 512        & 1      &  True          & LeakyReLU  \\
Conv       & 1          & 1      &  False         & None \\
\bottomrule
\end{tabular}%
}
\vspace{-1em}
\end{table}

\begin{table}[h]
\caption{CycleGAN generator architecture~\cite{CycleGAN2017}. Each row is a Convolution-InstanceNorm-ReLU layer.}
\vspace{-1em}
\label{tab:generator}
\resizebox{\columnwidth}{!}{%
\begin{tabular}{ccccc}
\toprule
Layer Name                                        &  Filters & Kernel & Stride & Padding \\
\midrule
c7s1-64                                                 & 64     & 7$\times$7    & 1      & Valid          \\
d128                                                     & 128    & 3$\times$3    & 2      & Same           \\
d256                                                     & 256    & 3$\times$3    & 2      & Same           \\
\multirow{2}{*}{R256 (9 times)}        & 256    & 3$\times$3    & 1      & Valid         \\
                                                         & 256    & 3$\times$3    & 1      & Valid         \\
u128                                                     & 128    & 3$\times$3    & 1/2    & Valid        \\
u64                                                      & 64     & 3$\times$3    & 1/2    & Valid         \\
c7s1-3                                                   & 3      & 7$\times$7    & 1      & Valid     \\
\bottomrule
\end{tabular}%
}
\vspace{-1em}
\end{table}

\end{document}